\documentclass[11pt]{article}

\usepackage{fullpage}
\usepackage{amsmath}
\usepackage{amssymb}
\usepackage{setspace}
\usepackage{bbm}
\usepackage{dsfont}
\usepackage{graphics}

\usepackage{hyperref}

\hypersetup{
    bookmarks=true,%
    colorlinks,%
    citecolor=blue,%
    filecolor=blue,%
    linkcolor=blue,%
    urlcolor=blue
}

\usepackage{color}

\usepackage[normalem]{ulem}

\newcommand{\be}{\begin{equation}}
\newcommand{\ee}{\end{equation}}
\newcommand{\bes}{\begin{equation*}}
\newcommand{\ees}{\end{equation*}}
\newcommand{\bea}{\begin{eqnarray}}
\newcommand{\eea}{\end{eqnarray}}
\newcommand{\beas}{\begin{eqnarray*}}
\newcommand{\eeas}{\end{eqnarray*}}

\newcommand{\bmat}{\begin{bmatrix}}
\newcommand{\emat}{\end{bmatrix}}

\newcommand{\RR}{\mathbb{R}}

\newcommand{\ZZ}{\mathbb{Z}}

\def\Tr{{\rm Tr}}
\def\le{\left}
\def\ri{\right}

\def\t{\tau}

\def\Tr{{\rm Tr}}
\def\le{\left}
\def\ri{\right}

\newcommand{\R}{\mathbb{R}}
\newcommand{\Z}{\mathbb{Z}}

\begin{document}
\numberwithin{equation}{section}
{
\begin{titlepage}
\begin{center}

\hfill \\
\hfill \\
\vskip 0.6in

{\Large \bf Warped Weyl fermion partition functions}\\
\vskip 0.4in

{\large  Alejandra Castro${}^{a}$, Diego M. Hofman${}^{a}$ and G\'abor S\'arosi${}^{b,c}$}\\

\vskip 0.3in

${}^{a}${\it Institute for Theoretical Physics, University of Amsterdam,
Science Park 904, Postbus 94485, 1090 GL Amsterdam, The Netherlands} \vskip .5mm
${}^{b}${\it Department of Theoretical Physics, Institute of Physics, Budapest University of Technology, H-1521 Budapest, Hungary } \vskip .5mm
${}^{c}${\it Kavli Institute for Theoretical Physics, Kohn Hall, University of California, Santa Barbara CA 93106-4030, USA} \vskip .5mm
{\it E-mail:} \href{mailto:a.castro@uva.nl}{a.castro@uva.nl}, \href{mailto:d.m.hofman@uva.nl}{d.m.hofman@uva.nl}, \href{mailto:sarosi@phy.bme.hu}{sarosi@phy.bme.hu}

\end{center}

\vskip 0.45in

\begin{center} {\bf ABSTRACT } \end{center}

Warped conformal field theories (WCFTs) are a novel class of non-relativistic theories. A simple, yet non-trivial, example of such theory is a massive Weyl fermion in $(1+1)$-dimensions, which we study in detail. We derive general properties of the spectrum and modular properties of partition functions of WCFTs. The periodic (Ramond) sector of this fermionic system is non-trivial, and we build two novel partition functions for this sector which have no counterpart in a CFT$_2$.  The thermodynamical properties  of WCFTs are revisited in the canonical and micro-canonical ensemble. 

%

\end{titlepage}
}

\newpage

\tableofcontents
\newpage 
\section{Introduction}

Conformal field theories (CFTs) represent the building blocks of our knowledge of quantum field theory. This is particularly true when it comes to strongly coupled theories. They constitute the low energy limit of all relativistic quantum field theories that are not gapped and, as such, they universally describe physical systems of interest.  As relevant as their  physical applications are the mathematical properties of CFTs; these properties allow for remarkable progress in their precise understanding. One line of research where the power of CFTs has become manifest is the bootstrap program \cite{Migdal:1972tk,Polyakov:1974gs, Ferrara:1973yt, Rattazzi:2008pe}. It has been understood that the analytic properties of four point functions in CFTs contain basic information that constrain the theory severely. It is even possible that minimal data might be used to fully determine the complete landscape of non-trivial CFTs. This would be a huge triumph of theoretical physics and might solve some important puzzles like the nature of the elusive (2,0) maximally supersymmetric CFT in six dimensions \cite{Witten:2007ct}.

An even more powerful case is that of two dimensional CFTs. It is well known that here the conformal group is upgraded to two sets of infinite dimensional Virasoro algebras \cite{Ginsparg:1988ui}. This allows for great mathematical control of these theories. It is possible, for example, to understand the general properties of thermal states and finite volume vacua from a simple minded analysis of the quantum theory on the two dimensional euclidean plane. Furthermore, the bootstrap program mentioned above can be fully carried out for two dimensional CFTs with central charge $c<1$ and, in this case, the landscape of unitary CFTs is completely understood \cite{yellowpages}. 

There is one extra feature of two dimensional CFTs which gives them great power and mathematical structure. This is the concept of modular invariance. CFTs defined on the torus have very interesting mathematical properties which strongly constrain the form of the partition function. Modular invariance is the statement that a CFT can be quantized on any of the two non trivial cycles of the torus and the resulting partition function should not be affected. While one can't prove this feature from first principles --it is necessary to add it as an extra postulate-- this is a very natural property that is suggested from the path integral formulation of the quantum theory. A striking consequence of modular invariance is that the asymptotic density of states at high energies is fully determined by the central charges. This is known as the Cardy formula \cite{Cardy:1986ie}.  It is natural to wonder: are there  other quantum field theories with a similar behavior?

Interestingly, it has proven very hard to extend the concept of modular invariance to more general types of quantum field theories. See, however, \cite{DiPietro:2014bca,Shaghoulian:2015kta,Basar:2015xda} for a discussion in CFTs in more than two dimensions and \cite{Gonzalez:2011nz,Shaghoulian:2015dwa} for some results in non relativistic field theories. Because of the great power of modular invariance, it would be of importance to understand larger families of quantum field theories where such a concept is available. 

There is of course one more crucial reason why CFTs have played such an important role in theoretical physics; namely, the AdS/CFT correspondence \cite{Maldacena:1997re,Witten:1998qj,Gubser:1998bc}. It has been long understood that quantum theories of gravity on Anti de Sitter (AdS) space-times are dual to CFTs in one less dimension. One of the main reasons to suspect the existence of such holographic principle is the fact that the entropy of black holes scales with their area and not their volume as one might have expected. This fact is, however,  very universal and not connected directly to the peculiarities of AdS space-times. The obvious question then becomes: how do we understand holography in non-AdS space-times?

Actually the questions raised above are heavily interconnected. One important clue in this direction was the observation that, in many holographic setups, the entropy of black holes seemed to present a Cardy-like behavior if understood as accounting for the asymptotic density of states of dual quantum field theories. The most interesting cases are given by (near) extremal rotating Kerr black holes \cite{Guica:2008mu,Castro:2009jf} and warped AdS$_3$ space-times \cite{Anninos:2008fx, Anninos:2013nja}. Interestingly, these spaces do not posses the local $SL(2,\R) \times SL(2, \R)$ isometry group typical of AdS$_3$ holography which implies the Cardy scaling. It was understood in \cite{Detournay:2012pc} that a class of two dimensional non-relativistic theories indeed has enough structure:  Warped Conformal Field Theories (WCFTs), first described in \cite{Hofman:2011zj},  can  account for this Cardy scaling and, moreover, enjoy the global $SL(2,\R) \times U(1)$ symmetry group manifest in the holographic setups mentioned above. WCFTs exhibit an infinite dimensional Virasoro-Kac-Moody $u(1)$ algebra. As such they are in principle as powerful as traditional two dimensional CFTs. It was later understood in \cite{Hofman:2014loa} what is a natural holographic description of WCFTs.

Therefore, we have candidate quantum field theories of a new class with enough mathematical structure that have well defined properties under modular transformations (necessary to explain the holographic Cardy scaling) and are connected to well understood non-AdS holographic setup. This set of theories is no longer mysterious. In \cite{Hofman:2014loa}, explicit free examples were constructed exploiting the understanding of the underlying non-relativistic geometric structure. One such example is the warped Weyl fermion, which we study in detail in the present work.

Having explicit examples of WCFTs and understanding them completely is of great importance both in field theory and holography. In quantum field theory little is known about non-relativistic  fixed points, with the exemption of a few very interesting examples such as  \cite{Cardy:1992tq}. Understanding exotic fixed points is difficult since is very easy to miss them if the scaling properties are unknown. As an example, warped Weyl fermions, while constituting a non-relativistic fixed point, flow to a gapped phase under usual relativistic scaling. In holography, our present knowledge strongly suggests WCFTs are critical in understanding the entropy of extremal black holes \cite{Detournay:2012pc}. We revisit the thermodynamic properties of WCFTs, and focus on the particular example of the warped Weyl fermion in this work. Understanding precisely how this connects to the problem of micro-state counting is an important problem on which we comment briefly further on.

An ambitious question we hope to address by studying the particular example at hand is: what is the meaning of modular invariance? As we explained above, this is not at all clear in the case of non-relativistic field theories. This question has important applications in real world physics. There are examples, like quantum Hall physics, where our example might be highly relevant \cite{Cappelli:1996np,Cappelli:2010be,Ryu:2012he}. In particular, in the case of chiral CFTs this question has been addressed in \cite{Ryu:2012he}. The question in this case is whether one can construct modular invariant partition functions without including an anti-holomorphic sector. The answer is affirmative. We present similar results for our warped Weyl fermion.

Lastly, we should comment on possible applications of this discussion to CFTs with a $\mathcal{W}_N$ extended algebra. In these theories we do not have a fully covariant formulation where higher spin currents and the energy momentum tensor are treated on equal footing. Furthermore, our notion of the renormalization group does not give special status to the higher conserved currents. Since the $\mathcal{W}_N$ algebra does not factorize we should aim to treat the whole symmetry structure democratically. The fact that we don't know how to do this is related to our lack of understanding of the properties of partition functions of these theories with higher spin chemical potentials turned on beyond perturbation theory \cite{Gaberdiel:2012yb,deBoer:2013gz,deBoer:2014fra}. WCFTs are an example of theories where, in this case, a spin one current is treated on equal footing with a conserved spin two current \cite{Hofman:2014loa}. Therefore, understanding the modular properties of partition functions in explicit WCFT examples might shed light on the more complicated $\mathcal{W}_N$ case.

The structure of this paper is as follows. In section \ref{sec:plane} we describe the canonical quantization of a (free) warped Weyl fermion on a cylinder. Similarly to what is done in CFTs we will exploit the mappings from the plane to the cylinder to define the theory in radial quantization.

In section \ref{sec:ontorus} we consider the theory on the torus. We define the partition function for this class of theories and study the transformation properties of such quantity with respect to changing the space circle where states are defined with respect to the the preferred axis of WCFTs. We calculate explicitly these functions for the warped Weyl fermion. Furthermore we explain the meaning of modular invariance in these theories and construct invariant partition functions by combining fermions with different boundary conditions, similarly to the way done in CFTs. In doing so we obtain novel combinations not considered in the usual CFT case. Also, we discuss explicitly the primary field content of the warped Weyl fermion.

In section \ref{sec:thermo} we discuss briefly the thermodynamic properties of WCFTs in general and discuss in particular the case of the warped Weyl fermion. We show that the entropy transforms covariantly under the allowed changes of coordinates in a WCFT. Furthermore, we elucidate the ``high temperature" limit of the entropy, shedding light on previous results discussed in  \cite{Detournay:2012pc}. While the results are straight forward in the canonical ensemble, we stress some subtleties that are encountered when one tries to extend these results to the micro-canonical ensemble. The possible implications of these details for the applications of WCFT formulae in holography are briefly mentioned.

Lastly we end with a short discussion and a few appendices including technical details and the necessary background to make the discussion in this paper self contained. Appendix \ref{app:wcft} contains background material on WCFTs. Appendix \ref{app:theta} contains our conventions for modular theta functions. Lastly, appendix \ref{app:fude} presents a derivation of warped partition functions from functional determinants.

\section{Warped Weyl fermion: from the plane to the cylinder}\label{sec:plane}

In this section we discuss how to quantize a WCFT. We first focus on a specific fermionic system which is an unitary example that includes explicitly a deformation that breaks the CFT symmetries. For this reason, this example  highlights unique features related to the system being non-relativistic.\footnote{An even simpler example would be a chiral CFT, since one could organize a chiral theory  using WCFT language --rather than usual CFT variables. However this class of theories lacks the space-time interpretation of symmetries which are intrinsic to a WCFT.}  In the last portion of this section we discuss general properties of the quantization procedure that are quantitatively distinct from a CFT.

%
To start we gather some basic definitions related to WCFTs. The defining feature of a WCFT is its symmetries:  provided two directions $(x^+, x^-)$, the system is invariant under the following transformations
\be\label{a1}
x^{+}~ \to~ x^+ + g(x^-)  ~,\quad  x^- ~\to ~f(x^-) ~,
\ee
where $f$ and $g$ are arbitrary functions. It is clear that both directions are not on the same footing: $x^-$ has scaling symmetry while $x^+$ does not. It is also evident that a WCFT is non-relativistic.  The global isometry group of the theory is $SL(2,\RR)\times U(1)$, while the local symmetries are described by a Virasoro-Kac-Moody algebra. Further properties of this algebra are described in appendix \ref{app:wcft}.

 A WCFT possesses  two global charges associated to energy and angular momentum:
\be\label{chargecyl}
H= - i \partial_t = - i (\partial_+ + \partial_-)~, \quad J= - i \partial_\varphi  = -i (\partial_+ - \partial_-) ~,
\ee
where 
\be\label{eq:cyl}
x^- + x^+ =  2 t ~, \quad x^+ - x^- = 2 \varphi~.
\ee

We want to define states in this theory at $t=0$ by doing radial quantization in the complex plane and gluing these states to the Lorentzian cylinder \eqref{eq:cyl}. For this purpose it is convenient to define the following complex coordinate
\be\label{zcoor}
z= e^{-i x^-} = e^{-i (t-\varphi)} = e^{t_E+ i\varphi}~, 
\ee
\noindent where $t_E=- i t$ corresponds to euclidean time. We have chosen $z$ such that it contains all time evolution. Very early time in the Euclidean cylinder corresponds to the origin in the $z$ plane. Now we define a second coordinate that does not involve time, namely:
\be\label{wcoor}
w = x^+ - x^- = 2\varphi~.
\ee
It is important that both (\ref{zcoor}) and (\ref{wcoor}) are included in the finite coordinates transformations generated by \eqref{a1}. They are therefore --up to anomalies-- symmetries of our theory.

In terms of the zero modes of the Virasoro-Kac-Moody defined on the cylinder, we can write 
\be\label{chargecyl2}
H= P_0^{\rm cyl} + L_0^{\rm cyl}~, \quad J= P_0^{\rm cyl} -L_0^{\rm cyl}~.
\ee
On the other hand, in the plane coordinates $(z,w)$ our charges (\ref{chargecyl}) are given, up to anomalies, by
\be
H = - z \partial_z ~,\quad J= - 2 \, i \partial_w +  z \partial_z~.
\ee
These differential operators correspond to charges defined on the plane by usual radial quantization. We can write them in terms of plane Virasoro-Kac-Moody charges and include as well the contribution from anomalies. The transformation we  perform  is a
standard map from the cylinder to the plane combined with a spectral flow transformation, with spectral flow parameter $\gamma=-\frac{1}{2}$ . Using \eqref{app:finite}, the exact relations are
\be\label{eq:hj2}
H = L_0 - \frac{c}{24} - \frac{k}{4} ~,\quad J = 2 P_0 - L_0 + \frac{c}{24} - \frac{3}{4} k ~,
\ee
\noindent where $k$ is the Kac-Moody level and $c$ the Virasoro central charge.

\subsection{Warped Weyl Fermion}\label{sec:qww}

A free Warped Weyl fermion \cite{Hofman:2014loa} is a complex anti-commuting field $\Psi$ whose dynamics is described by the action 
\be
I= \int dtd\varphi \le(i\bar \Psi\partial_+ \Psi + m\bar\Psi \Psi\ri)~.
\ee  
In particular note that scaling invariance $x^-\to \lambda x^-$ is preserved, provided $\Psi$ transforms as a weight ${1\over 2}$ operator under warped conformal transformations \cite{Hofman:2014loa}. This example, while simple, has  no CFT counterpart due to the mass term.   This introduces new features as we quantize the theory by using radial quantization. 

Our starting point is to understand the on-shell properties of the fermion. It's equations of motion are given by 
\be
\partial_+ \Psi - i m \Psi =0~, \quad \partial_+ \bar \Psi + i m \bar \Psi =0~. 
\ee
The solutions to these equations are given by 
\be
\Psi(x^+,x^-) = e^{i m x^+} \psi(x^-)~, \quad \bar \Psi (x^+,x^-)= e^{-i m x^+} \bar \psi(x^-)~,
\ee
with arbitrary Grassmann functions $\psi$ and $\bar \psi$. However, the field should have a definite periodicity on the cylinder. Given that we set  $\varphi \sim \varphi +2\pi$, we can demand periodic boundary condition, which we denote by ``R'', and in this case we must have
 \be\label{eq:r1}
{\rm R}: \quad \psi(x^- - 2\pi ) = e^{-  2 \pi i\, m} \psi(x^-) ~, \quad \bar \psi(x^- - 2\pi ) = e^{ 2 \pi i \, m} \bar\psi(x^-)~.
\ee
If we instead impose anti-periodic (NS) boundary  conditions, we need
 \be\label{eq:ns1}
{\rm NS}:\quad \psi(x^- - 2\pi ) =- e^{-  2 \pi i\, m} \psi(x^-) ~, \quad \bar \psi(x^- - 2\pi ) = -e^{  2 \pi i \, m} \bar\psi(x^-)~.
\ee

Taking into account that $\psi(x^-)$ and $\bar \psi(x^-)$   transform as a weight $\frac{1}{2}$ operator, the fields $(\Psi, \bar \Psi)$ in terms of the coordinates $(z, w)$ are
\be
\Psi= e^{i m w} z^{-m}  \left( i z^{-1} \right)^{\frac{1}{2}} \psi(z)~,\quad \bar \Psi= e^{-i m w} z^{m}  \left( i z^{-1} \right)^{\frac{1}{2}} \bar \psi(z)~.
\ee
Looking at the above expression, it is convenient to introduce a pair of plane fields $(\eta,\bar \eta)$ such that
\be
\Psi= e^{i m w} \eta(z) ~,\quad \bar\Psi= e^{-i m w} \bar \eta(z)~.
\ee
In this notation, the boundary conditions \eqref{eq:r1} and \eqref{eq:ns1} now read
 \be
{\rm R}: \quad \eta (e^{2 \pi i} z) = e^{ 2\pi i (-2 m - \frac{1}{2}) } \eta(z) ~, \quad  \bar \eta (e^{2 \pi i} z) = e^{ 2\pi i (2 m - \frac{1}{2}) } \bar\eta(z)~,
\ee
and
 \be
{\rm NS}:\quad  \eta (e^{2 \pi i} z) = -e^{ 2\pi i (-2 m - \frac{1}{2}) } \eta(z) ~,\quad  \bar \eta (e^{2 \pi i} z) = -e^{ 2\pi i (2 m - \frac{1}{2}) } \bar\eta(z)~.
\ee
This means we can write the following mode expansion for the $(\eta, \bar \eta)$ fields
\be
\eta(z) = \sum_n z^{-n -2m -\frac{1}{2}} \eta_n ~, \quad \bar \eta(z) = \sum_n z^{-n +2m -\frac{1}{2}} \bar \eta_n ~,
\ee
\noindent where the sum is over integers in the R sector and semi-integers in the  NS sector. The fact that $\Psi$ and $\bar \Psi$ are complex conjugate fields implies that
\be
\eta_n^\dag = \bar \eta_{-n}~,
\ee
and canonical quantization dictates that 
\be
\{ \bar \eta_n, \eta_{n'} \} = \delta_{n+n'}~.
\ee
Using these mode anti-commutators one can obtain the following OPE:
\be\label{eq:ope}
\bar \eta(z) \eta(z') \sim \frac{1}{z-z'}~.
\ee

For this particular field theory we can define two holomorphic conserved currents related to $x^-$ and $x^+$ translations respectively. Namely,
\be\label{eq:tp1}
T(z) = -\frac{1}{2} \left( \bar \eta \partial_z \eta + \eta \partial_z \bar \eta\right) ~,\quad P(z) = m \, \bar \eta\, \eta~.
\ee
The OPE \eqref{eq:ope} then implies 
\begin{eqnarray}
T(z) T(z') &\sim& \frac{1/2}{(z-z')^4} + \frac{2}{(z-z')^2} T(z') + \frac{1}{z-z'} \partial_{z'} T(z')~,\cr
T(z) P(z') &\sim& \frac{1}{(z-z')^2} P(z') + \frac{1}{z-z'} \partial_{z'}P(z')~,\cr
P(z) P(z') &\sim& \frac{m^2}{(z-z')^2}~,
\end{eqnarray}
which ensures the canonical Virasoro-Kac-Moody algebra at $c=1$ and $k= 2 m^2$. 

We can now define vacua on the plane by demanding that the $\eta, \bar \eta$ fields do not create large singularities at the origin. We would like our original field in the cylinder to be regular once analytically continued onto the plane. This requires:
\begin{eqnarray}\label{eq:reg}
&&\eta_n | 0\rangle = 0 \quad \textrm{if} \quad n+2 m + \frac{1}{2} > \frac{1}{2} ~,\cr
&&\cr
&&\bar \eta_n | 0\rangle = 0 \quad \textrm{if} \quad n-2 m + \frac{1}{2} > \frac{1}{2}~.
\end{eqnarray}
These requirements fix the two point functions on the plane to be
\be\label{eq:cf1}
\langle \bar \eta(z) \eta(z') \rangle = \frac{1}{\sqrt{z z'}} \sum_{\ell \ge 2 m} \left(\frac{z'}{z}\right)^{\ell - 2m}~.
\ee
The sum is over $\ell$ integers in the R sector and semi-integers in the NS sector. Writing these sums is not hard if we define
\be
 k =2m^2 \equiv \frac{1}{2} \left(Q-\alpha\right)^2~,
\ee
\noindent where $Q$ is an integer and $\alpha \in [0,1)$. 
The correlation function \eqref{eq:cf1} is then
\be\label{eq:bnn}
\langle \bar \eta(z) \eta(z') \rangle = \left\{ \begin{array}{cc} 
\left(\frac{z'}{z}\right)^{\alpha} \frac{1}{z-z'} &\quad \textrm{ in NS sector with} \quad \alpha \in [0,\frac{1}{2}) \\
\left(\frac{z'}{z}\right)^{\alpha-\frac{1}{2}} \frac{1}{z-z'} & \quad \textrm{ in R sector} \\
\left(\frac{z'}{z}\right)^{\alpha-1} \frac{1}{z-z'} & \quad \textrm{ in NS sector with} \quad \alpha \in [\frac{1}{2},1)
\end{array} \right.
\ee
By reverting the order in the correlator we also obtain
\be
\langle \eta(z) \bar\eta(z') \rangle = \left\{ \begin{array}{cc} 
\left(\frac{z}{z'}\right)^{\alpha} \frac{1}{z-z'} &\quad \textrm{ in NS sector with} \quad \alpha \in [0,\frac{1}{2}) \\
\left(\frac{z}{z'}\right)^{\alpha-\frac{1}{2}} \frac{1}{z-z'} & \quad \textrm{ in R sector} \\
\left(\frac{z}{z'}\right)^{\alpha-1} \frac{1}{z-z'} & \quad \textrm{ in NS sector with} \quad \alpha \in [\frac{1}{2},1)
\end{array} \right.
\ee

From $\langle \eta(z) \bar\eta(z') \rangle$ it is simple to calculate the vacuum values of our $P(z)$ and $T(z)$ operators. All we need to do is to take the limit of coincident points $z \to z'$ and remove the singular piece. The first finite term determines the expectation values. From \eqref{eq:tp1}, and using \eqref{eq:bnn}, 
\be
\langle P(z') \rangle =  \left\{ \begin{array}{cc} 
 -  \frac{m \alpha}{z'} & \quad  \textrm{ in NS sector with} \quad \alpha \in [0,\frac{1}{2}) \\
 -  \frac{m \left(\alpha-\frac{1}{2}\right)}{z'} & \quad \textrm{ in R sector} \\
 -  \frac{m \left(\alpha-1\right)}{z'} &  \quad \textrm{ in NS sector with} \quad \alpha \in [\frac{1}{2},1)
\end{array} \right.
\ee
and from here we obtain
\be\label{eq:po1}
\langle P_0 \rangle =  \left\{ \begin{array}{cc} 
 -  m \alpha& \quad  \textrm{ in NS sector with} \quad \alpha \in [0,\frac{1}{2}) \\
 -  m \left(\alpha-\frac{1}{2}\right) & \quad \textrm{ in R sector} \\
 -  m \left(\alpha-1\right) &  \quad \textrm{ in NS sector with} \quad \alpha \in [\frac{1}{2},1)
\end{array} \right.
\ee
Similarly for the Virasoro current, we have
\be
\langle T(z') \rangle =  \left\{ \begin{array}{cc} 
 \frac{\alpha^2}{2 z'^2} & \quad  \textrm{ in NS sector with} \quad \alpha \in [0,\frac{1}{2}) \\
   \frac{\left(\alpha-\frac{1}{2}\right)^2}{2 z'^2} & \quad \textrm{ in R sector} \\
  \frac{\left(\alpha-1\right)^2}{ 2 z'^2} &  \quad \textrm{ in NS sector with} \quad \alpha \in [\frac{1}{2},1)
\end{array} \right.
\ee
\noindent which implies
\be\label{eq:lo1}
\langle L_0 \rangle =  \left\{ \begin{array}{cc} 
  \frac{\alpha^2}{2} & \quad  \textrm{ in NS sector with} \quad \alpha \in [0,\frac{1}{2}) \\
  \frac{\left(\alpha-\frac{1}{2}\right)^2}{2} & \quad \textrm{ in R sector} \\
   \frac{\left(\alpha-1\right)^2}{2} &  \quad \textrm{ in NS sector with} \quad \alpha \in [\frac{1}{2},1)
\end{array} \right.
\ee

One interesting consequence of \eqref{eq:po1} and \eqref{eq:lo1} is that the vacuum state we have defined indeed minimizes the hamiltonian of the system $H$. In order to show this it suffices to consider states that are Kac-Moody primaries. For these states we have
\be
L_0 = \frac{P_0^2}{k} = \frac{P_0^2}{2 m^2} ~.
\ee
Furthermore, the spectrum of $P_0$ is quantized in units of $m$ up to a shift given by the vacuum energy calculated above. Explicitly, we have primary states defined by integers $q$,
\be
P_0 | q \rangle = \left(m q + \langle P_0 \rangle\right) | q \rangle~.
\ee
This is a direct consequence of our fields $(\bar \eta, \eta)$ having charges $(m, -m)$ respectively. Therefore
\be\label{eq:loq}
\langle q | L_0 | q \rangle =   \left\{ \begin{array}{cc} 
  \frac{\left(q - \alpha\right)^2}{2} & \quad  \textrm{ in NS sector with} \quad \alpha \in [0,\frac{1}{2}) \\
  \frac{\left(q- \alpha+\frac{1}{2}\right)^2}{2} & \quad \textrm{ in R sector} \\
   \frac{\left(q-\alpha+1\right)^2}{2} &  \quad \textrm{ in NS sector with} \quad \alpha \in [\frac{1}{2},1)
\end{array} \right.
\ee
For both R and NS boundary conditions, $L_0$ is minimal for $q=0$ in agreement with \eqref{eq:lo1}. Using this result, the value of $H$ on the vacuum state for both sectors is
\be
\langle H \rangle =  \left\{ \begin{array}{cc} 
  \frac{\left(\alpha\right)^2}{2} -\frac{c}{24} - \frac{k}{4}& \quad  \textrm{ in NS sector with} \quad \alpha \in [0,\frac{1}{2}) \\
  \frac{\left(\alpha-\frac{1}{2}\right)^2}{2}  -\frac{c}{24} - \frac{k}{4}& \quad \textrm{ in R sector} \\
   \frac{\left(\alpha-1\right)^2}{2}  -\frac{c}{24} - \frac{k}{4}&  \quad \textrm{ in NS sector with} \quad \alpha \in [\frac{1}{2},1)
\end{array} \right.
\ee
\noindent where $c=1$ and $k=2m^2$ in our case. The global vacuum lies in the R sector for $\alpha \in (\frac{1}{4},\frac{3}{4})$ and in the NS sector otherwise. The angular momentum at $q=0$ is
\be\label{eq:jvac}
\langle J \rangle =  \left\{ \begin{array}{cc} 
  -\frac{1}{2} \left( \alpha + 2m\right)^2+\frac{c}{24}+\frac{k}{4} & \quad  \textrm{ in NS sector with} \quad \alpha \in [0,\frac{1}{2}) \\
   -\frac{1}{2} \left( \alpha -\frac{1}{2}+ 2m\right)^2  +\frac{c}{24}+\frac{k}{4}& \quad \textrm{ in R sector} \\
  -\frac{1}{2} \left( \alpha -1+ 2m\right)^2  +\frac{c}{24}+\frac{k}{4}&  \quad \textrm{ in NS sector with} \quad \alpha \in [\frac{1}{2},1)
\end{array} \right.
\ee
Notice that $J$ could be arbitrarily negative by considering descendant states with large $L_0$. The value of $J$ is however bounded above. It is not hard to see that only a primary could maximize $J$. The value of $J$ for a primary labeled by $q$ is:
\be
\langle q | J | q \rangle =  \left\{ \begin{array}{cc} 
  -\frac{1}{2} \left( Q -q\right)^2 +\frac{c}{24} +\frac{k}{4}& \quad  \textrm{ in NS sector with} \quad \alpha \in [0,\frac{1}{2}) \\
   -\frac{1}{2} \left( Q -\frac{1}{2}- q\right)^2 +\frac{c}{24}+\frac{k}{4}& \quad \textrm{ in R sector} \\
  -\frac{1}{2} \left(Q-1- q\right)^2  +\frac{c}{24}+\frac{k}{4}&  \quad \textrm{ in NS sector with} \quad \alpha \in [\frac{1}{2},1)
\end{array} \right.
\ee
where we used that $\alpha +2m$ must be an integer which we defined to be $Q$.
 Therefore  the state $| Q \rangle$ maximizes the angular momentum in the first case, the state $| Q-1 \rangle$ in the third and both states are degenerate for the middle case. The maximum value for the angular momentum is
\be\label{eq:jmaxf}
 J_{\rm max} =  \left\{ \begin{array}{cc} 
\frac{c}{24} +\frac{k}{4} & \quad  \textrm{ in NS sector with} \quad \alpha \in [0,\frac{1}{2}) \\
\frac{c}{24}+\frac{k}{4}-\frac{1}{8}& \quad \textrm{ in R sector} \\
\frac{c}{24}+\frac{k}{4}&  \quad \textrm{ in NS sector with} \quad \alpha \in [\frac{1}{2},1)
\end{array} \right.
\ee

Let us now comment on the angular momentum quantization conditions. Say the vacuum angular momentum is quantized (we will come back to this question shortly). If that is the case, it is comforting to see that all other primaries and descendants are (half-integer) quantized as well. This is because we can write
\be\label{}
\langle q | J | q\rangle = \langle 0 | J | 0\rangle + \left\{ \begin{array}{cc} 
Q q - \frac{q^2}{2}  & \quad  \textrm{ in NS sector with} \quad \alpha \in [0,\frac{1}{2}) \\
  Q q -\frac{q}{2}- \frac{q^2}{2}& \quad \textrm{ in R sector} \\
Q q -q- \frac{q^2}{2}&  \quad \textrm{ in NS sector with} \quad \alpha \in [\frac{1}{2},1)
\end{array} \right.
\ee
\noindent where $Q$ is as before the integer we use to define $2m$. So if say $\langle 0 | J | 0\rangle$ is half integer, then so is $\langle q | J | q\rangle$. Descendants can only add integers to these values. 

 To close this discussion, the last consistency check is that $\langle 0 | J | 0\rangle$ is indeed quantized in all sectors of the theory. Consider first the case where our candidate theory contains both the R and NS sectors: here we will need to impose an additional condition that the jump in angular momentum is quantized as well. Allowing for half integer quantization, from \eqref{eq:jvac} we must have
\be
\frac{Q}{2}-\frac{1}{8} \in \frac{\mathbb{Z}}{2}~.
\ee
However this is not possible! Not having a quantized  spectrum for $J$  has catastrophic consequences for modular invariance as the $T$ transformation (see (\ref{Ttran}) further on) depends on this fact. One option is to live with this and demand a weaker version of modular invariance, e.g. that it applies only to $T^4$. Another route is to consider a system consisting of several copies of complex fermions. This is not unfamiliar from chiral massless fermions, see e.g. \cite{Ryu2012Interacting}. If we have $N$ complex fermions we need to rescale the values of our ground state charges as
\be
\langle L_0 \rangle \rightarrow N \langle L_0 \rangle ~,\quad \langle P_0 \rangle \rightarrow N \langle P_0 \rangle~.
\ee
This implies in our theory of free fermions\footnote{More generally, we could have $c=N+c_0$ where $c_0$ corresponds to the central charge coming from neutral (under $P_0$) degrees of freedom, if there where any. For a free Warped Weyl fermion $c_0=0$.}
\be\label{inch}
c= N ~,\quad k =2 N m^2~.
\ee
\noindent 
The quantization condition for the difference of angular momentum between vacua becomes
\be\label{condns}
\frac{N Q}{2}-\frac{N}{8} \in \frac{\mathbb{Z}}{2} \rightarrow N = 4 n \quad \textrm{with} \quad n \in \mathbb{Z}~.
\ee

Finally, we as well need that each individual vacuum has a quantized angular momentum. For the R sector this implies
\be\label{rcond}
6 k + c - 3 N = 12 p \quad \textrm{with} \quad p \in \mathbb{Z}~.
\ee
If we use \eqref{condns}, this just becomes 
\be
6 k + c  = 12 p' \quad \textrm{with} \quad p' \in \mathbb{Z}~,
\ee
\noindent where we have redefined $p$. Using $c=N=4n$, we have
\be\label{eq:cnsj}
k = 2p - {N\over 6} =2p - \frac{2}{3} n~.
\ee
Therefore, the smallest theories contain 4 complex fermions and $k= \frac{4}{3}, \frac{10}{3}, \frac{16}{3},\ldots$.

It might also be possible to construct a theory that only contains the R sector (see section \ref{sec:ontorus}). If this is the case, the quantization constraints are relaxed and only (\ref{rcond}) must be imposed, which for $N$ complex fermions reads as
\be\label{eq:crj}
k={N\over 3} + 2p ~, \quad p\in \ZZ~.
\ee
In particular, there exist theories satisfying these constraint with just one complex fermion and $k= 2 p +\frac{1}{3} = \frac{1}{3}, \frac{7}{3}, \frac{13}{3},\ldots$.

\subsection{General Properties of Radial Quantization}\label{sec:2grq}

One rather interesting and surprising feature of quantizing the warped Weyl fermion is the behavior of angular momentum. In particular we found in \eqref{eq:jmaxf} an upper bound for  $J$. This has no counterpart in a CFT. Nevertheless, we now show that this feature is general for any WCFT. To see this explicitly, consider again  $J$ expressed in term of plane charges \eqref{eq:hj2}:
\be
J = 2 P_0 - L_0 + \frac{c}{24} - \frac{3}{4} k ~.
\ee
We can remove the dependence on $P_0$ by simply doing a spectral flow transformation. Using \eqref{eq:spectralflow} with $\gamma=1$ gives
\be\label{eq:ja}
J = -L_0^{(\gamma=1)} + \frac{c}{24}+{k\over 4}~.
\ee
The spectrum of $L_0^{(\gamma=1)}$ is bounded from below by the unitarity bounds discussed in appendix \ref{app:wcft}, and hence 
\be
J\leq J_{\rm max} \equiv  \frac{c}{24}+{k\over 4}~.
\ee
The important distinction, is that in a CFT one does not have access to spectral flow transformations that implement spacetime rotations. This is the key to write $J$ in terms of a bounded operator. 

We can as well in a simple manner derive consequences of quantizing the spectrum of $J$. From \eqref{eq:ja} this will obviously require that $L_0^{(\gamma=1)}$ has integer spacing. Furthermore, if we denote by $h$ the lowest eigenvalue of  $L_0^{(\gamma=1)}$ we have
\be
h+ \frac{c}{24}+{k\over 4} \in {\Z\over 2}~,
\ee
where we are allowing for half integer quantization. This is in agreement with the constraints \eqref{rcond} since $h$ is an integer for the fermion. 


It is useful to make another comparison with CFTs. If we blindly follow CFT lore, we would argue that the existence of the identity operator is correlated with requiring that the vacuum state on the plane satisfies $\langle L_0 \rangle=0$. This argument clearly fails for a WCFT. The reason is that the map used leaves out a dependence on $\varphi$ through the variable $w$ which is considered fixed on the $z$ plane. This means that regularity conditions on the plane do not translate to the appropriate quantization conditions on the cylinder. This is clear, since in \eqref{eq:reg} we demanded regularity which turned in to $\langle L_0 \rangle \neq 0$ in  \eqref{eq:lo1}. What is universally true is that for the vacuum state the unitarity bound is saturated, i.e.
\be\label{eq:lower}
\langle L_0 \rangle = \frac{\langle P_0^2 \rangle}{k}~.
\ee 
These primaries are as well preserved by the global $SL(2,\RR)\times U(1)$ symmetries of the theory.

To end, consider one last time \eqref{eq:hj2}; in an arbitrary spectral flow frame we will have
\bea
H &=& L_0^{(\gamma)}+2\gamma P_0^{(\gamma)} - \frac{c}{24} - \frac{k}{4}\le(1-4\gamma^2\ri) ~,\cr
J &=& 2(1-\gamma) P_0^{(\gamma)} - L_0^{(\gamma)} + \frac{c}{24} - \frac{3}{4} k + k\gamma(2-\gamma)~.
\eea
The lowest energy state  is a Kac-Moody primary which saturates the unitarity bound, i.e. $h_\gamma={p_\gamma^2\over k}$. For these states we find
\bea
H &=& {1\over k}\le(p_\gamma +\gamma k\ri)^2 - \frac{c}{24} - \frac{k}{4}~,\cr
J &=& -{1\over k}\le(p_\gamma +(\gamma-1) k\ri)^2  + \frac{c}{24} + \frac{k}{4} ~.
\eea
The appealing feature of this formula is that we isolated the dependence on the spectral flow parameters relative to the anomalies. If we casted the anomalous piece in CFT language, we would find
\be\label{eq:crcl}
c_R= c + 6k~, \quad c_L=0~.
\ee
This identification is obviously ambiguous; the only appealing feature is that $c_R$ is the combination that dictates quantization condition of $J$. 


\section{On to the Torus}\label{sec:ontorus}

In this section we construct partition functions for the free fermion system in section \ref{sec:qww}. Our aim is to highlight properties and consequences of modular invariance in WCFTs, and use the fermion as  an explicit realization of these properties.

As our starting point, we first review how warped symmetries act on the torus. Consider a theory compactified on the $\varphi$ circle with period $2\pi$ as in the previous section. Let us put this theory at temperature $\beta^{-1}$ and angular potential $\mu$. In this frame our identifications are
\be\label{torid}
(x^- , x^+) \sim (x^- - 2\pi, x^+ + 2\pi) \sim (x^- - 2\pi \tau, x^+ + 2 \pi \bar{\tau})~,
\ee
\noindent where as before
\be
x^+ = t+ \varphi ~,\quad x^-=t-\varphi~,
\ee
\noindent and we have defined
\be
2 \pi \tau = \mu - i \beta~,\quad 2\pi \bar \tau =  \mu + i \beta~.
\ee
The first identification in (\ref{torid}) corresponds to the spatial identification, while the second one is thermal. The modular group is described by two transformations: $S$ which corresponds to interchanging thermal and spatial cycles; and $T$ which accounts for adding the spatial cycle to the thermal cycle.

However, we could have chosen our cycles rather differently. This will be important  for what follows, so let us consider a rather arbitrary torus defined by the following identifications
\be\label{eq:uv1}
(u , v) \sim (u - 2\pi \ell, v + 2\pi \bar \ell) \sim (u - 2\pi \tau, v + 2 \pi \bar{\tau})~.
\ee
In this new parametrization, $u$ is the scaling direction while $v$ defines the $U(1)$ axis. The partition function is, thus,\footnote{It is important to note that $L_0^{\rm cyl}$ and $P_0^{\rm cyl}$ in \eqref{eq:z1} are the generators of translations in the $(u,v)$ directions; we are using the same notation as in \eqref{chargecyl2} to avoid clutter, and to distinguish them from the plane charges. The choice of torus, and hence the relevant zero modes, will be made explicit by the subscript $(\ell,\bar \ell)$ in the partition function. }
\be\label{eq:z1}
Z_{\bar \ell| \ell} (\bar \tau| \tau)= \Tr_{\bar\ell, \ell} \le(e^{2 \pi i \bar \tau P_0^{\rm cyl}} e^{- 2\pi i \tau L_0^{\rm cyl}}\ri)~.
\ee
 We can now use a change of coordinates to connect this partition function to a canonical one where $u$ defines the spatial identification with period $2\pi$. This can be done by the change of coordinates:
\be\label{eq:huv}
\hat u = \frac{u}{\ell} ~,\quad \hat v=v +\frac{\bar \ell}{\ell} u~.
\ee
Keeping track of the anomalies, we obtain
\be\label{spectral}
Z_{\bar \ell |  \ell} (\bar \tau| \tau) = e^{\pi i k \bar \ell \left( \bar \tau -  \frac{\tau \bar \ell}{2 \ell} \right)} Z_{0|1} (\bar \tau - \frac{\bar \ell \tau}{\ell}| \frac{\tau}{\ell}) \equiv  e^{\pi i k \bar \ell \left( \bar \tau -  \frac{\tau \bar \ell}{2 \ell} \right)}  \hat Z(\bar \tau - \frac{\bar \ell \tau}{\ell} | \frac{\tau}{\ell})~,
\ee
\noindent where we defined the function $\hat Z(\cdot | \cdot)$ by the equation above. In relation to our choice of spatial cycle in section \ref{sec:plane} we have
\be\label{conc}
Z_{1|1} (\bar \tau | \tau) = e^{\pi i k  \left( \bar \tau -  \frac{\tau}{2} \right)}  \hat Z(\bar \tau - \tau | \tau)~.
\ee

 It is the function $\hat Z$ that has nice modular properties so we can just calculate this partition function and obtain all other possible ones from the formulae above. This is evident by the following argument: under the exchange of the spatial and thermal circles we expect
\be\label{eq:zs01}
Z_{0|1} (\bar \tau|\tau) = Z_{\bar \tau| \tau} (0|-1) ~,
\ee
which defines $S$ invariance on the torus. We will come back to \eqref{eq:zs01} in subsection \ref{sec:gmi}; for now we declare it as a symmetry of the system. Then, using  \eqref{spectral} on $Z_{\bar \tau | \tau} (0|-1) $, we arrive at\footnote{In this section $z=\bar\tau-\tau$, with $\ell= \bar \ell=1$. It should not be confused with the complex coordinate used in section \eqref{zcoor}. We hope the context makes clear the definition of $z$.}
\be\label{eq:shz}
 \hat Z (z |\tau) =e^{\pi i k \frac{z^2}{2 \tau}} \hat Z (\frac{z}{\tau} | -\frac{1}{\tau})~.
\ee
This implies that under $S$ the partition function $Z_{1,1}$ transforms as 
\bea
Z_{1|1} (\bar \tau| \tau) &=&  e^{\pi i k \frac{\bar \tau^2}{2 \tau}} \hat Z (\frac{\bar \tau - \tau}{\tau} | -\frac{1}{\tau})\cr
 &=&  e^{\pi i k \frac{(\bar \tau-1)^2}{2 \tau}} Z_{1|1} (\frac{\bar \tau - \tau-1}{\tau} | -\frac{1}{\tau})~.
\eea
Clearly the transformation properties of  $Z_{1,1}$ are rather unnatural relative to  $\hat Z$; this makes the frame $(\bar \ell,\ell)=(0,1)$ preferred.

Invariance under $T$ is tied to quantization of angular momentum along the compact spatial direction, and hence (anti-)periodicity of our fields along that direction. Say we quantized the system as in section \ref{sec:plane}; then the partition function should satisfy\footnote{One could of course relax \eqref{eq:t1} such that the partition is invariant under, for example, $T^2$. This would amount to quantizing $J$ in half integer units. } 
\be\label{eq:t1}
Z_{1| 1}(\bar\tau +1 |\tau +1)= Z_{1| 1}(\bar\tau|\tau )~,
\ee
and from \eqref{conc} this implies that
\be\label{eq:thz}
 \hat Z(z | \tau) = e^{\pi i {k \over 2 }} \hat Z(z | \tau +1)~.
\ee

The derivations presented so far are in complete agreement with \cite{Detournay:2012pc}. What is highlighted here, and will be important in the following, is that relations such as \eqref{eq:shz} are dependent of the choice of axis used to parametrize the torus. $\hat Z(z|\tau)$ is preferred from this point of view relative to, for example,  $Z_{1|1}(\bar\tau,\tau)$.

\subsection{Warped Weyl Fermion}\label{sec:torusweyl}

The task ahead of us is to classify possible $\hat Z$'s built from characters of the R and NS sectors for the free Warped Weyl fermion. We will demand that these partition functions transform according to \eqref{eq:shz} under $S$ along with appropriate invariance under $T$. 

Written as a trace, the partition  function $\hat Z$ expressed in terms of plane generators $\hat L_0$, $\hat P_0$ reads
\be
\hat Z (z | \tau) = \Tr \le( e^{2\pi i z \hat P_0}{e^{-2\pi i \tau (\hat L_0 - \frac{c}{24})}}\ri)~.
\ee
Notice that these coordinates are privileged as the vacuum is generated by the identity operators,  i.e.  the vacuum state is neutral under $P_0$. We can use the following formulae to connect with the generators of the previous section:
\be\label{eq:lp3}
\hat L_0 = L_0 - 2 P_0 + k ~,\quad \hat P_0 = P_0 - k~.
\ee
 While there is a one to one map between the states in different spatial circles, the vacuum does not map to itself. Therefore the state with lowest $L_0$ depends on this choice. For example, the vacuum in this preferred coordinates $(\hat u,\hat v)$ does not correspond to the vacuum of our spatial circle in section \ref{sec:qww}. 
Instead, the state with maximal angular momentum is the vacuum! 

More explicitly, in the NS sector we have
\bea\label{vac:ns}
&\hat L_0 | Q \rangle =0~,\quad \hat P_0| Q \rangle =0  ~,\quad  &\textrm{if} \quad \alpha \in [0,\frac{1}{2})~, \cr
&\hat L_0 | Q-1 \rangle =0 ~,\quad \hat P_0| Q-1 \rangle =0 ~,\quad&  \textrm{if} \quad \alpha \in [\frac{1}{2},1)~.
\eea
where we used \eqref{eq:loq} and \eqref{eq:lp3}, and for the R sector 
\bea\label{vac:r}
&\hat L_0 | Q \rangle =\frac{1}{8}  | Q \rangle ~,\quad \hat P_0| Q \rangle =\frac{m}{2} |Q \rangle ~,\quad&  \textrm{if} \quad \alpha \in [0,\frac{1}{2}) ~,\cr
&\hat L_0 | Q -1\rangle =\frac{1}{8}  | Q-1 \rangle ~,\quad \hat P_0| Q-1 \rangle =-\frac{m}{2} |Q-1 \rangle ~,\quad & \textrm{if} \quad  \alpha \in [\frac{1}{2},1)~.
\eea
We see that without loss of generality we can keep $\alpha \in [0,\frac{1}{2})$ as these cases are interchangeable up to a change of sign of $\hat P_0$. We assume this from now on.

Then, we can define a new vacuum of our theory. In the NS sector
\be
| \hat 0 \rangle_{\rm NS} =  \bar \eta_{\frac{1}{2}} \ldots \bar \eta_{Q+\frac{1}{2}} | 0 \rangle_{\rm NS}~.
\ee
In the R sector
\be
| \hat - \rangle_{\rm R} = \bar \eta_1  \ldots \bar \eta_{Q-1} | 0 \rangle_{\rm R} \quad\quad | \hat + \rangle_{\rm R} = \bar \eta_0  \ldots \bar \eta_{Q-1} | 0 \rangle_{\rm R} ~.
\ee 
Notice that these states do not minimize the Hamiltonian in the last section! They will be useful however to calculate the partition function explicitly.

Now let us calculate $\hat Z$ for our free fermion system. We introduce different boundary conditions in the thermal circle by including (or not) the operator
\be
(-1)^F = e^{-\pi i \frac{\hat P_0}{m}}~.
\ee
So we define the usual partition functions
\bea
&&\hat Z_{\rm RR} = q^{-\frac{1}{24}} \Tr_{\rm R}\le( (-1)^F y^{\frac{\hat P_0}{m}} q^{\hat L_0}\ri)~, \quad \hat Z_{\rm RNS} = q^{-\frac{1}{24}} \Tr_{\rm R}  \le(y^{\frac{\hat P_0}{m}} q^{\hat L_0}\ri) ~,\cr
&&\cr
&&\hat Z_{\rm NSNS} = q^{-\frac{1}{24}} \Tr_{\rm NS}\le( y^{\frac{\hat P_0}{m}} q^{\hat L_0}\ri) ~,\quad 
\hat Z_{\rm NSR} = q^{-\frac{1}{24}} \Tr_{\rm NS}\le( (-1)^F y^{\frac{\hat P_0}{m}} q^{\hat L_0}\ri)~,
\eea
\noindent where we defined
\be
y =e^{2\pi i m z} ~,\quad q= e^{-2\pi i \tau}~.
\ee
Note that inserting $(-1)^F$ amounts to $y \rightarrow -y$. We can now calculate these functions by creating all states with fermionic oscillators:\footnote{The fact that the theta functions have a minus signs is due to our conventions \eqref{torid}-\eqref{eq:z1}. These minus signs have no physical repercussions.}
\bea\label{eq:zis}
\hat Z_{\rm RR} (z | \tau)&=& -i y^{\frac{1}{2}} q^{\frac{1}{12}} \prod_{n \ge 0} \left(1- q^{n+1} y \right) \left(1- q^n y^{-1}\right)=\frac{\theta_1 ( m z | - \tau)}{\eta(-\tau)}~,\cr
\hat Z_{\rm RNS}(z | \tau) &=& y^{\frac{1}{2}} q^{\frac{1}{12}} \prod_{n \ge 0} \left(1+q^{n+1} y \right) \left(1+q^n y^{-1}\right)=\frac{\theta_2 ( m z | - \tau)}{\eta(-\tau)}~,\cr
\hat Z_{\rm NSNS} (z | \tau)&=&  q^{-\frac{1}{24}} \prod_{n \ge 0} \left(1+q^{n+\frac{1}{2}} y \right) \left(1+q^{n+\frac{1}{2}} y^{-1}\right)= \frac{\theta_3 ( m z | - \tau)}{\eta(-\tau)}~,\cr
\hat Z_{\rm NSR}(z | \tau) &=&  q^{-\frac{1}{24}} \prod_{n \ge 0} \left(1-q^{n+\frac{1}{2}} y \right) \left(1-q^{n+\frac{1}{2}} y^{-1}\right)= \frac{\theta_4 ( m z | - \tau)}{\eta(-\tau)}~.
\eea
The prefactors of $q$ and $y$ account for the vacuum values in \eqref{vac:ns} and \eqref{vac:r}; the factor of $i$ in $\hat Z_{\rm RR}$ is just for ease of writing it in terms of known modular forms. Our conventions on theta functions and Dedekind's eta function are in appendix \ref{app:theta}.

With these explicit expression for the characters, it is straight forward to extract their transformation properties under $S$ and $T$.
We can use the Jacobi identities \eqref{eq:jacobi} to write
\bea\label{eq:sc}
\hat Z_{\rm RR}(z | \tau)& =&   i e^{i \pi m^2 \frac{z^2}{\tau}}   \hat Z_{\rm RR}(\frac{z}{\tau} | -\frac{1}{\tau}) ~,\cr
\hat Z_{\rm RNS}(z | \tau) &= &   e^{i \pi m^2 \frac{z^2}{\tau}}   \hat Z_{\rm NSR}(\frac{z}{\tau} | -\frac{1}{\tau})~, \cr
\hat Z_{\rm NSNS}(z | \tau) &=&    e^{i \pi m^2 \frac{z^2}{\tau}}   \hat Z_{\rm NSNS}(\frac{z}{\tau} | -\frac{1}{\tau})~,\cr
\hat Z_{\rm NSR}(z | \tau) &= &   e^{i \pi m^2 \frac{z^2}{\tau}}   \hat Z_{\rm RNS}(\frac{z}{\tau} | -\frac{1}{\tau}) ~.
\eea
This should be compared with the $S$ transformation \eqref{eq:shz}. Recall that $m^2=\frac{k}{2}$, hence the phase factors are those capturing the anomalous term in $S$!
However, the individual characters do not satisfy \eqref{eq:shz}. For instance, the factor of $i$ spoils $S$ covariance for $\hat Z_{\rm RR}$, unless we consider two complex fermions --allowing invariance in the double cover-- or 4 complex fermions for full invariance.
On the other hand it seems perfectly possible to make $S$ invariant combinations of single fermions by combining the other 3. An $S$ invariant expression is
\be\label{eq:guess}
 \alpha\,  \hat Z_{\rm NSNS}(z | \tau)  + \beta \left(\hat Z_{\rm RNS}(z | \tau) + \hat Z_{\rm NSR}(z | \tau) \right)~,
\ee
with $\alpha$ and $\beta$ suitable real parameters. However, as we will see shortly, this is not compatible with $T$ transformations.

For a CFT, invariance under $T$ is not very restrictive, if not automatic in many cases. In a WCFT, while we impose the same physical requirement, it does limit the system more dramatically.  The $T$ transformation comes from momentum quantization on the circle of interest. For the Weyl fermion we chose the frame $(\ell,\bar\ell)=(1,1)$, and hence quantization of $J$ is equivalent to invariance under 
\be \label{Ttran}
T : \tau \rightarrow \tau +1 ~,\quad \bar \tau \rightarrow \bar \tau +1~.
\ee
Using the properties of the theta functions \eqref{eq:Tshifts} gives
\bea\label{eq:tc}
\hat Z_{\rm RR}(z | \tau+1) &=& e^{-i \frac{\pi}{6}} \hat Z_{\rm RR}(z | \tau) ~,\cr
\hat Z_{\rm RNS}(z | \tau+1) &=& e^{-i \frac{\pi}{6}} \hat Z_{\rm RNS}(z | \tau) ~,\cr
\hat Z_{\rm NSNS}(z | \tau+1) &=& e^{i \frac{\pi}{12}} \hat Z_{\rm NSR}(z | \tau) ~,\cr
\hat Z_{\rm NSR}(z | \tau+1) &=& e^{i \frac{\pi}{12}} \hat Z_{\rm NSNS}(z | \tau) ~.
\eea
In a CFT these phases are present but irrelevant, since the partition functions are composed of complex conjugate pairs of characters. For a WCFT this is no longer the case, which gives rise to further restrictions.  

It is now clear that we can construct two inequivalent modular invariant partition functions, which satisfy \eqref{eq:shz} and \eqref{eq:thz} simultaneously. Starting with $\hat Z_{\rm RR}$, for full invariance under $S$ we need to consider $N=4 n$ complex fermions, and hence
\be\label{pfr}
\hat Z_{\rm R, full} (z | \tau) = \left[ \hat Z_{\rm RR}(\frac{z}{\sqrt{4 n}} | \tau)\right]^{4 n} ~, \quad n\in \Z~.
\ee
Demanding $T$ invariance restricts the $U(1)$ level to be
\be\label{kr}
k = \frac{4 n}{3}  + 4 p~, \quad  p\in \Z~.
\ee
This reproduces \eqref{eq:crj} up to a factor of 2 simply because we allowed for semi-integer quantization in section \ref{sec:qww}. While only imposing $T$ invariance allows to have just $N=1$ in the R sector, $S$ invariance forces us to consider only multiple of four complex fermions. There is no counterpart of \eqref{pfr} in a CFT, since for a CFT the relevant holomorphic character is $\hat Z_{\rm RR} (z=0 | \tau) $ which is trivial. The partition function for the R sector in a WCFT is non-trivial, and this makes \eqref{pfr} unique and novel. 

The other modular invariant combination comes by combining characters of the NS sector. The combination in \eqref{eq:guess} is  compatible with $S$ but clearly not uniform with respect to $T$ due to \eqref{eq:tc}. In this case, invariance under $T$ requires that we have  a multiple of 8 complex fermions --such that the powers of the characters transforms uniformly-- and then it furthers restricts $k$ by demanding \eqref{eq:thz}. Hence, a modular invariant partition function in the NS sector is
\be\label{pfns}
\hat Z_{\rm NS,full} (z | \tau) =  \left[\hat Z_{\rm NSNS}(\frac{z}{\sqrt{8 n}} | \tau)\right]^{8n}  + \left[\hat Z_{\rm RNS}(\frac{z}{\sqrt{8 n}} | \tau)\right]^{8 n} + \left[\hat Z_{\rm NSR}(\frac{z}{\sqrt{8 n}} | \tau)\right]^{8 n}~, 
\ee
with $c=N=8n$ and $n\in \Z$. This combination transforms as required by $S$, and $T$ invariance further requires
\be
k=-\frac{4n}{3} + 4 p = \frac{8n}{3}  + 4 p'~,
\ee
\noindent for some integer $p$ or $p'$. This agrees again with our previous results \eqref{eq:cnsj} if we demand integer quantization. Notice this is the same as (\ref{kr}) in terms of the total number of fermions $N$.

\subsection{Primary content of the Warped Weyl fermion theory}\label{sec:prim}

As it stands, the modular invariant partition functions \eqref{pfr} and \eqref{pfns} are casted in terms of fermionic respresentations of the Virasoro-Kac-Moody algebra.  We now decompose these partitions functions in terms of bosonic characters of the algebra. This makes evident that in, e.g., \eqref{pfns} all coefficients in the sum are positive, and hence might hint to an interesting process of bosonization in WCFTs. 

More concretely, we want to organize the partition function\footnote{We could of course discuss this decomposition for $Z_{\bar \ell|\ell}(\bar \tau|\tau)$ as well; for sake of simplicity we focus on $\hat Z (z| \tau)$. }
\be\label{eq:x1}
\hat Z (z| \tau)= \Tr \le(e^{2 \pi i z  \hat P_0^{\rm cyl}} e^{- 2\pi i \tau \hat L_0^{\rm cyl}}\ri)
\ee
 in terms of primaries states of the algebra and its descendants. For this purpose, it is convenient to cast the spectrum of the Virasoro sector as
 \be
 \hat L_0 = \hat h+ {p^2\over k} ~,\quad\quad  \hat h \geq 0~,
 \ee
where we are accounting for the Sugawara contribution of $\hat P_0 = p$ to the Virasoro current; this just follows from \eqref{eq:lsug}. Here we will use $\hat h$ and $p$ as the quantum numbers of a primary state $|\hat h,p\rangle $ on the plane. Then \eqref{eq:x1} can be organized as
\be\label{eq:x2}
\hat Z (z| \tau)= q^{-c/24} \sum_{p}\,   \, e^{2\pi i z p} {q^{ p^2\over k} \over \phi(q)} \sum_{\hat h}  \chi_{\hat h}(q) q^{\hat h} ~,
\ee  
Here the sums run over values of $\hat h$ and $p$ in the spectrum; note that $\hat h$ could still have residual dependence on $p$, so the ordering of the summation is a priori non-trivial. The descendants created by acting with $P_{-n}$'s on   $|\hat h,p\rangle $ are accounted by the Euler phi function
\be
\phi(q)=\prod_{n=1}^{\infty}(1-q^n)~,
\ee
while the descendants arising from the action of $L_{-n}$'s are counted by an ordinary Virasoro character with central charge $c-1$: $\chi_{\hat h}(q)$.  

Our aim is to decompose the partition functions for the Warped Weyl fermion system as in \eqref{eq:x2}. To start  we consider again \eqref{eq:zis}, now expressed as a sum; the structure of each fermionic character can be summarized as   follows
\bea\label{eq:x3}
\hat Z^{(ab)}(z | \tau)= q^{-1/24} \sum_{r\in \mathbb{Z} +a/2} (-1)^{b r}y^{r} {q^{{r^2\over 2}}\over \phi(q)}~,
\eea
The parameters $(a,b)$ control boundary conditions around the spatial and temporal cycle respectively; for example, $a=0$ and $b=1$ corresponds to $\hat Z_{\rm NS R}$. \eqref{eq:x3} already resembles the structure in \eqref{eq:x2} with $\hat h=0$, $p=m r$, $k=2m^2$ and $\chi_{\hat h}(q)=1$ since a single fermion has $c=1$. 

However, the combinations of interest are those in \eqref{pfr} and \eqref{pfns} which involve powers of $\hat Z^{(ab)}$. In this case, we have that the $N$-th power of \eqref{eq:x3} is
\bea\label{eq:x4}
q^{-{N\over 24}}\sum_{\hat r\in \mathbb{Z} +{aN\over 2}  }(-1)^{b \hat r} y^{\hat r\over \sqrt{N}} {q^{{\hat r^2\over 2N}}\over \phi(q)} \left(\sum_{\{r_i\}\in  \mathbb{Z} +{a\over2} |\sum_i r_i=\hat r}\frac{q^{{1\over 2}(r_1^2+\ldots+r_N^2)-\frac{(r_1+\ldots+r_N)^2}{2N}}}{\phi(q)^{N-1}} \right)
\eea
In comparison to  \eqref{eq:x2},   we need to show that the term in parentheses can be interpreted as the sum of $N-1$  bosonic characters with \textit{non-negative} primary weights, i.e. a legitimate unitary Virasoro character. Indeed, we may write the weights as
\be
{r_1^2+...+r_N^2}-\frac{(r_1+...+r_N)^2}{N}=\sum_{i,i'=1}^N r_{i}(\delta_{ii'}-\frac{1}{N})r_{i'}~.
\ee
The symmetric matrix $A_{ii'}=(\delta_{ii'}-\frac{1}{N})$ has one vanishing eigenvalue corresponding to the total charge: the vector with all of its components being 1. Moreover, the $N-1$ dimensional subspace of vectors with vanishing sum of their components is an eigenspace of $A$ with eigenvalue 1. This clearly shows that $A$ is positive semidefinite. Let $O$ be the \textit{real} orthogonal matrix diagonalizing $A$: $OAO^T=\text{diag}(0,1,1,...,1)$. Then we may write
\be
\sum_{i,i'=1}^N r_{i}(\delta_{ii'}-\frac{1}{N})r_{i'}=\sum_{j=2}^N(\sum_{i=1}^NO_{ji}r_i)^2~.
\ee
Define $N-1$ weights associated to a charge vector $\vec{r}=(r_1,\ldots,r_N)$ as
\be
\hat h_j(\vec{r})={1\over 2}(\sum_{i=1}^NO_{ji}r_i)^2 \geq 0~, \quad \quad j=2,\ldots,N~,
\ee
which allows us to write
\be
\chi_{a,\hat{r}}(q)\equiv \sum_{\{r_i\}\in  \mathbb{Z} +{a\over 2} |\sum_i r_i=\hat r}\frac{q^{{1\over 2}(r_1^2+\ldots+r_N^2)-\frac{(r_1+\ldots+r_N)^2}{2N}}}{\phi(q)^{N-1}}=\sum_{\{r_i\}\in  \mathbb{Z} +{a\over2} |\sum_i r_i=\hat r} \,\prod_{j=2}^N\left( \frac{q^{\hat h_j(\vec{r})}}{\phi(q)} \right)~.
\ee
This is a product of $N-1$ scalar characters with positive weights, as claimed. Note that this expression depends only on $\hat n$, and the spatial boundary conditions through the choice of  $a=0,1$ in the sum. Replacing in \eqref{eq:x4} we have
\be\label{eq:x5}
\le(\hat Z^{(ab)}({z\over\sqrt{N}} | \tau)\ri)^N = q^{-{N\over 24}}\sum_{\hat r\in \mathbb{Z} +{aN\over 2}  }(-1)^{b \hat r} y^{\hat r\over \sqrt{N}} {q^{{\hat r^2\over 2N}}\over \phi(q)} \chi_{a,\hat{r}}(q)~.
\ee

We are now ready to evaluate the modular invariant functions for the fermion system. From \eqref{pfns} we have
\bea
\hat Z_{\rm NS,full} (z | \tau)&=& \le(\hat Z_{\rm NSNS}({z\over\sqrt{8n}} | \tau)\ri)^{8n}+ \le(\hat Z_{\rm RNS}({z\over\sqrt{8n}} | \tau)\ri)^{8n}+ \le(\hat Z_{\rm NSR}({z\over\sqrt{8n}} | \tau)\ri)^{8n}\cr
&=&2q^{-n/3}\sum_{\hat r\in 2\mathbb{Z}  }y^{\hat r\over \sqrt{8n}} {q^{{\hat r^2\over 16n}}\over \phi(q)} \chi_{0,\hat{r}}(q) +q^{-n/3}\sum_{\hat r\in \mathbb{Z}  } y^{\hat r\over \sqrt{8n}} {q^{{\hat r^2\over 16n}}\over \phi(q)} \chi_{1,\hat{r}}(q) ~.
\eea
While the individual contributions in \eqref{eq:x5} have negative coefficients, the modular invariant combination $\hat Z_{\rm NS,full} $ has only positive coefficients. 

The other modular invariant  $\hat Z_{\rm R,full} $ in \eqref{pfr} is of the form \eqref{eq:x5} and hence will have both positive and negative coefficients.\footnote{If we interpreted $\hat Z_{\rm R,full} $ as an index instead of a partition function it would be perfectly natural to have both positive and negative coefficients. Regardless, all coefficients are integral in $\hat Z_{\rm R,full} $.} However, it is possible to form combinations that involve the Ramond sector while keeping all coefficients positive. In particular, we have
\bea
 &&\le(\hat Z_{\rm NSNS}({z\over\sqrt{8n}} | \tau)\ri)^{8n}+ \le(\hat Z_{\rm RNS}({z\over\sqrt{8n}} | \tau)\ri)^{8n}+ \le(\hat Z_{\rm NSR}({z\over\sqrt{8n}} | \tau)\ri)^{8n} +  \le(\hat Z_{\rm RR}({z\over\sqrt{8n}} | \tau)\ri)^{8n} \cr
 &&\cr
&& =2q^{-n/3}\sum_{\hat r\in 2\mathbb{Z}  }y^{\hat r\over \sqrt{8n}} {q^{{\hat r^2\over 16n}}\over \phi(q)} \left(\chi_{0,\hat{r}}(q)  +\chi_{1,\hat{r}}(q)\ri) ~,
\eea
and 
\bea
 &&\le(\hat Z_{\rm NSNS}({z\over\sqrt{8n}} | \tau)\ri)^{8n}+ \le(\hat Z_{\rm RNS}({z\over\sqrt{8n}} | \tau)\ri)^{8n}+ \le(\hat Z_{\rm NSR}({z\over\sqrt{8n}} | \tau)\ri)^{8n} - \le(\hat Z_{\rm RR}({z\over\sqrt{8n}} | \tau)\ri)^{8n} \cr
 &&
 \cr
 &&=2q^{-n/3}\sum_{\hat r\in 2\mathbb{Z}  }y^{\hat r\over \sqrt{8n}} {q^{{\hat r^2\over 16n}}\over \phi(q)} \chi_{0,\hat{r}}(q)  +2q^{-n/3}\sum_{\hat r\in 2\mathbb{Z}+1  }y^{\hat r\over \sqrt{8n}} {q^{{\hat r^2\over 16n}}\over \phi(q)} \chi_{1,\hat{r}}(q)~. 
\eea
These two last expressions are rather unique to WCFTs; there is no counterpart of these combinations in a regular CFT. They also hint to a rather novel version of bosonization in WCFT. It would be interesting to explore if these combinations could be realized by a bosonic field on a lattice.

\subsection{On modular invariance for trace partition functions}\label{sec:gmi}

An important principle we have used throughout is invariance under $S$.  Around \eqref{eq:shz}  we used that the partition function that has a trivial modular invariant exchange property is $ Z_{0| 1} (\bar \tau| \tau)$. In particular we declared as a symmetry of the system that 
\be\label{modi1}
Z_{0| 1} (\bar \tau | \tau) = Z_{\bar \tau| \tau} (0|-1)~.
\ee
We can use this expression together with (\ref{spectral}) to construct the general exchange formula
\be\label{modi2}
Z_{\bar \ell| \ell} (\bar \tau| \tau) =  e^{\pi i k  \bar \tau \bar \ell} Z_{\bar \tau| \tau} (-\bar \ell| -\ell)~.
\ee
The sign change is necessary so we define the partition function where it is manifestly convergent.  Along these lines, the generalization of $T$ invariance is
\be
Z_{\bar \ell| \ell}(\bar\tau +\bar \ell |\tau +\ell)= Z_{\bar \ell| \ell}(\bar\tau|\tau )~,
\ee
provided we quantized angular momentum using the spatial cycle $(\bar\ell,\ell)$. From here one can infer that the generalization of \eqref{eq:thz} is
\be
 \hat Z(z | \tau) = e^{\pi i k {\bar \ell^2\over 2 }} \hat Z(z | \tau +1)~.
\ee

It is not unexpected that \eqref{modi2} has a anomalous piece, since the symmetry is anomalous. However, the exponent is not a priori evident: a different exponent would modify \eqref{modi1} and spoil our analysis for the fermion. In the following we would like to explain why our conditions are natural and physically motivated. 

One way to do so is by writing the partition function in a language that makes manifest the warped geometry. The symmetry is anomalous, so the  partition function could depend on quantities that are not invariant under warped conformal transformations. A good starting point is then to list the invariant combinations. Define vectors
\be
\vec p = (\bar p, p) ~, \quad \vec w = (\bar w, w) ~,
\ee
which are defined in the  $(v,u)$ plane. In this notation $u$ is the scaling direction in the warped geometry, and under this scaling a vector transforms as 
\be\label{eq:scale}
\vec p=(\bar p, p ) ~\rightarrow ~(\bar p , \lambda p)~.
\ee
There are two combinations that are invariant under \eqref{eq:scale} \cite{Hofman:2014loa}:
\be\label{eq:inv1}
\frac{ \vec p \times \vec w }{|\vec p|^\frac{1}{2} |\vec w|^\frac{1}{2}} ~, \quad \frac{|\vec p|}{|\vec w|}~,
\ee
where
\be
 |\vec p| \equiv  p ~, \quad  \vec p \times \vec w  \equiv p\bar w - \bar p w~. 
\ee
Notice that $|\vec p|$ is not necessarily positive. The combinations \eqref{eq:inv1} are moreover invariant under  warped conformal transformations; we refer the reader to \cite{Hofman:2014loa} for details and derivations. 

In this language, we notice that $\hat Z$ defines a quantity which is invariant under warped conformal transformations:
\be\label{eq:zinv}
\hat Z\left( \frac{\vec \ell \times \vec \tau}{|\vec \ell |^\frac{1}{2} |\vec \tau|^\frac{1}{2}} \frac{|\vec \tau|^\frac{1}{2}}{|\vec \ell|^\frac{1}{2}} \Big{|} \frac{|\vec \tau|}{|\vec \ell|} \right) =  e^{-\pi i k \bar \ell \left( \bar \tau - \frac{ \tau \bar \ell}{2 \ell} \right)} Z_{\bar \ell | \ell} \left( \bar \tau | \tau\right)~.
\ee
The phases involved in the expression above quantify exactly the breaking of the symmetry. 
Now we would like to define modular invariance in this setup. This is just the statement that partition functions are invariant under the exchange of the $\vec \ell$ and $\vec \tau$ cycles. This would amount to
\be
Z_{\bar \ell| \ell} (\bar \tau|\tau) \sim Z_{\bar \tau| \tau} (-\bar \ell| -\ell)~.
\ee
Now we show that the symbol $\sim$ above cannot mean equality as this would contradict the fact that the function $\hat Z$ is only a function of invariant quantities. Let us define
\be
Z_{\bar \ell | \ell} (\bar \tau | \tau) \equiv e^{f(\vec \ell, \vec \tau)} Z_{\bar \tau | \tau} (-\bar \ell |-\ell)~.
\ee
This is of course always possible for a suitable function $f$. Using this expression in \eqref{eq:zinv} we obtain
\be
\hat Z\left( \frac{\vec \ell \times \vec \tau}{|\vec \ell |^\frac{1}{2} |\vec \tau|^\frac{1}{2}} \frac{|\vec \tau|^\frac{1}{2}}{|\vec \ell|^\frac{1}{2}} \Big{|} \frac{|\vec \tau|}{|\vec \ell|} \right)  =e^{ f(\vec \ell, \vec \tau)- i \pi k \bar \tau \bar \ell} e^{\frac{\pi i k}{2} \frac{\left(\vec \ell \times \vec \tau\right)^2}{|\vec \tau| |\vec \ell|}} \hat Z\left( \frac{\vec \ell \times \vec \tau}{|\vec \ell |^\frac{1}{2} |\vec \tau|^\frac{1}{2}} \frac{|\vec \ell|^\frac{1}{2}}{|\vec \tau|^\frac{1}{2}} \Big{|} -\frac{|\vec \ell|}{|\vec \tau|} \right)~. 
\ee
The left hand side is an invariant quantity, therefore $f$ can't be trivial nor invariant under warped transformations. In order to cancel the non-invariant term in the right hand side, we must have
\be\label{modi3}
f(\vec \ell, \vec \tau) = i \pi k \bar \tau \bar \ell + g(\vec \ell, \vec \tau)~
\ee 
where $g(\vec \ell, \vec \tau)$ is an arbitrary function that is invariant under warped conformal transformations.
The natural definition of modular invariance in a WCFT is, thus, 
\be\label{modi4}
g(\cdot)=0 ~. 
\ee

For further reference we quote this result in covariant form
\be\label{stran}
\hat Z\left( \frac{\vec \ell \times \vec \tau}{|\vec \ell |}  \Big{|} \frac{|\vec \tau|}{|\vec \ell|} \right)  = e^{\frac{\pi i k}{2} \frac{\left(\vec \ell \times \vec \tau\right)^2}{|\vec \tau| |\vec \ell|}} \hat Z\left( \frac{\vec \ell \times \vec \tau}{ |\vec \tau|} \Big{|} -\frac{|\vec \ell|}{|\vec \tau|} \right)~,
\ee
and explicitly
\be\label{stran2}
\hat Z\left( \bar \tau - \bar \ell \frac{\tau}{\ell} \Big{|} \frac{\tau}{\ell} \right)  = e^{\frac{\pi i k}{2} \frac{\left( \ell \bar \tau - \bar \ell \tau\right)^2}{\tau \ell}} \hat Z\left( \frac{\ell}{\tau} \bar \tau - \bar \ell \Big{|} -\frac{ \ell}{\tau} \right)~. 
\ee

Notice that in \cite{Detournay:2012pc} the transformation of the partition function under the exchange of cycles was assumed to be (\ref{modi1}). Here we can establish that this is the case as a consequence of (\ref{modi3}) and (\ref{modi4}).


\section{Thermodynamics}\label{sec:thermo}

We will now discuss the thermodynamic properties of WCFTs. We will use modular invariant properties of the preferred frame $(\bar \ell | \ell) = (0|1)$ to extract  properties of the system at ``high temperatures.'' From here we can evaluate the entropy and other thermodynamic properties in any other frame.

These calculations are done in the canonical ensemble, and in this ensemble we will recover and extend the results in \cite{Detournay:2012pc}. It is also of interest to comment on the relation of these calculations to the micro-canonical counterpart. In concrete terms, we will discuss the consequences of modular invariance on the density of states of WCFTs.

\subsection{Canonical ensemble}

Given a partition function $Z$, it is possible to define a canonical entropy $S$ through
\be\label{entropy1}
S_{\bar \ell | \ell} (\bar \tau | \tau) \equiv \left(1 - \bar \tau \partial_{\bar \tau} - \tau \partial_\tau \right) \log Z_{\bar \ell | \ell} (\bar \tau | \tau) ~.
\ee
Notice that this quantity can only have statistical meaning as a canonical entropy whenever the thermodynamic potentials $\bar \tau$ and $\tau$ are purely imaginary. In that case $Z$ is a statistical partition function. If this was not the case, it would be some kind of index and  we could still calculate $S$, although we have to be careful in interpreting it physically.

Making use of (\ref{spectral}), we can easily obtain $S_{\bar \ell | \ell}$ in terms of the entropy defined in the preferred frame $( \bar \ell | \ell) = (0,1)$. We have
\begin{eqnarray}
S_{\bar \ell | \ell}(\bar \tau | \tau) &=& \left(1 - \bar \tau \partial_{\bar \tau} - \tau \partial_\tau \right) \log [e^{\pi i k \bar \ell \left( \bar \tau -  \frac{\tau \bar \ell}{2 \ell} \right)}  \hat Z(\bar \tau - \frac{\bar \ell \tau}{\ell} | \frac{\tau}{\ell}) ] \nonumber\\ 
&=&  \left(1 - z \partial_{z} - t \partial_t \right) \log \hat Z (z| t) \Big|_{z = \bar \tau - \bar \ell \frac{\tau}{\ell}\, , \, t = \frac{\tau}{\ell}}\nonumber\\
&=& \hat S (\bar \tau - \bar \ell \frac{\tau}{\ell} | \frac{\tau}{\ell}) ~.
\end{eqnarray}
The entropy, in terms of thermodynamic potentials, is blind to the anomalous transformation of $Z$. This is to be expected if one wants to give statistical meaning to $S$ and at the same time view (\ref{spectral}) as a coordinate transformation. All observers must agree on the results of the counting problem. Hence, without loss of generality, we will evaluate $\hat S$.

Now, using modular invariance we can calculate the entropy of WCFTs in the ``high temperature" regime, i.e. $\tau \rightarrow -i 0$, and recover the results in \cite{Detournay:2012pc}. We can use (\ref{stran2}) to write
\be\label{entropyhat}
\hat S(z | t) = \left(1 - z \partial_{z} - t \partial_t \right) \log  [ e^{ \frac{\pi i k}{2} \frac{z^2}{t}} \hat Z ( \frac{z}{t} | - \frac{1}{t}) ]=  \left(1 - z \partial_{z} - t \partial_t \right) \log   \hat Z ( \frac{z}{t} | - \frac{1}{t}) ~.
\ee
Once again, the anomalous factor drops out of the calculation. If we understand the $\tau \rightarrow - i 0$ limit of this expression we can calculate the entropy. Here we proceed as in standard CFT. We can argue that the partition function will be dominated by the most important term in the trace over states. In particular,
\be
\hat Z (\frac{z}{t} | -\frac{1}{t}) = \textrm{Tr} \left( \, e^{2 \pi i \frac{z}{t} \hat P_0 + 2 \pi i {1\over t} \left(\hat L_0 - \frac{c}{24} \right)} \right)\rightarrow e^{2 \pi i \frac{z}{t} \hat P^*_0 + 2 \pi i {1\over t} \left(\hat L^*_0 - \frac{c}{24} \right)} ~,
\ee
\noindent where $\hat L^*_0$ is the minimum value of $\hat L_0$ in the spectrum and $\hat P^*_0$ is the charge of that state. It is important to notice that this result is only strictly valid for imaginary values of $z$ and  when the spectrum of $\hat P_0$ is real up to an overall shift. If this was not the case the result above has to be revised and we will comment on it shortly. If we plug this in the expression for the entropy (\ref{entropyhat}) we obtain
\be
\hat S(z | t) = 2 \pi i \frac{z}{t} \hat P^*_0 + 4 \pi i {1\over t} \left(\hat L^*_0 - \frac{c}{24} \right) ~.
\ee
This is the result quoted in \cite{Detournay:2012pc}  for the preferred quantization circle $(\bar \ell | \ell) = (0 | 1)$. We notice that the first term is unusual, if compared to standard CFT results.

The minimal value of $L_0$ is given by primaries that saturate the unitarity bound; see \eqref{eq:lower}. In that case
\be\label{normalL02}
\hat L^*_0 =  \frac{\hat P^{*2}_0}{k} ~,
\ee
and we must find the state with lowest absolute value of $\hat P_0$. It is natural to assume that the spectrum contains a neutral state, which gives  $\hat P^*_0 =0$, and hence the entropy is\footnote{In general we cannot prove that a neutral state exists for any WCFT. However,  by using  \eqref{eq:shz} twice gives $ \hat Z (z |\tau)= \hat Z (-z |\tau)$; this further implies that the spectrum is invariant under $\hat P_0 \rightarrow - \hat P_0$. If the neutral state is not present, we would simply write a generalization of \eqref{eq:szt} that accounts for the reflection symmetry.}
\be\label{eq:szt}
\hat S(z | t) = - \pi i \frac{c}{6 t}  ~.
\ee
And in a generic frame the entropy is given by
\be\label{wentropyf}
S_{\bar \ell | \ell} (\bar \tau | \tau ) = -\pi i \ell \frac{c}{6 \tau} ~,
\ee
\noindent when $\frac{\tau}{\ell} \rightarrow - i 0$.  Notice that the entropy does not exhibit any dependence on the thermodynamic potential $\bar \tau$, nor the circle data $\bar \ell$.  These manipulations so far are valid for the Warped Weyl Fermion studied in previous sections. Therefore, the modular invariant combinations constructed in section \ref{sec:ontorus} for a system of $N$ Warped Weyl Fermions exhibit a canonical entropy given by \eqref{wentropyf} with $c=N$. 

It is instructive to recast the results above in terms of the expectation values of the operators $P_0$ and $L_0$ given the thermodynamic potentials. These are given by
\begin{eqnarray}
\langle P_0 \rangle &=& \frac{1}{2 \pi i} \partial_{\bar \tau} \log Z_{\bar \ell | \ell} (\bar \tau | \tau)~,\cr
\langle  L_0 \rangle &=& -\frac{1}{2 \pi i} \partial_{\tau} \log Z_{\bar \ell | \ell} (\bar \tau | \tau)~.
\end{eqnarray}
In this case the anomalous terms in the transformation contribute to the calculation. Keeping track of these terms we obtain:
\be
\langle P_0 \rangle 
=\frac{\ell}{\tau} \left(  \frac{k}{2} \bar \tau +\hat P^*_0 \right)~.
\ee
and 
\be
\langle L_0 \rangle
= \frac{\ell}{\tau^2} \left(\frac{k}{4} \bar \tau^2 + \bar \tau \hat P^*_0 + \hat L^*_0 - \frac{c}{24} \right)~.
\ee
For our Warped Weyl Fermion we can plug in $\hat P^*_0 = \hat L^*_0 = 0$ and obtain
\begin{eqnarray}\label{eq:41}
\langle P_0 \rangle &=& \frac{\ell}{\tau}\frac{k}{2} \bar \tau~,\cr
\langle L_0 \rangle&=&  \frac{\ell}{\tau^2} \left(\frac{k}{4} \bar \tau^2  - \frac{c}{24} \right) ~.
\end{eqnarray}
It is clear from the expressions above that an imaginary value for $\bar \tau$ yields arbitrary real values for $\langle P_0 \rangle$ and negative imaginary values for $\tau$ yields positive values for $\langle L_0 \rangle$. This justifies the interest in this limit where we can give a statistical meaning to $S$. 

Plugging \eqref{eq:41} in (\ref{wentropyf}) we recover, in any frame, the expected chiral CFT result:
\be
S_{\bar \ell | \ell} = 2 \pi \sqrt{ \frac{c}{6} \left( \langle L_0 \rangle - \frac{\langle P_0 \rangle^2}{k} \right)}~.
\ee
The fact that the entropy depends only on the spectral flow invariant quantity $ \langle L_0 \rangle - \frac{\langle P_0 \rangle^2}{k}$ is consistent with micro-canonical arguments that show that Virasoro-Kac-Moody representations with all charges have the same number of states at a given Virasoro level \cite{Carlip:1998qw}.

To make a more precise connection to the micro-canonical ensemble we will need to integrate the expressions above over the thermodynamic potentials $\bar \tau$ and $\tau$. We will discuss this in the next subsection. Therefore, we might be interested in the behavior of $Z$ in the complex $z$ plane. Notice that if $z$ is not purely imaginary the most dominant term in the trace could be different. The lowest states still satisfy (\ref{normalL02}), but this time the dominant term of the sum is given by the a different minimization problem:
\be
\hat P_0^* : \min{ \{\operatorname{Re}(z) \, \hat P_0 + \frac{\hat P_0^2}{k}\}} \rightarrow \hat P^*_0 = - \frac{k}{2} \operatorname{Re}(z) = - \frac{k}{2} \operatorname{Re}(\bar \tau) ~.
\ee
It is important to make clear that we are assuming the spectrum is dense enough that there exists a state with a charge in the neighborhood of this value. Still , if the spectrum is discrete we should stress that no differential operator involving $z$ acts non trivially on $\hat P^*_0$. We can therefore adapt our old formulae to this case.
In particular, the expectation values of the charges are now
\be
\langle P_0 \rangle =\frac{\ell}{\tau} \frac{k}{2} i \operatorname{Im}(\bar \tau) ~,
\ee
and 
\be
\langle L_0 \rangle = -\frac{\ell}{\tau^2} \left(\frac{k}{4} \left(\operatorname{Im}(\bar \tau)\right)^2 +\frac{c}{24} \right) ~.
\ee
We see they are independent of the real part of $\bar \tau$. As such, the expression for the entropy in this limit is necessarily changed to 
\be\label{eq:4s}
S_{\bar \ell | \ell} = 4 \pi i\frac{\hat P^*_0 \langle \hat P_0 \rangle}{k} +  2 \pi \sqrt{ \frac{c}{6} \left( \langle L_0 \rangle - \frac{\langle P_0 \rangle^2}{k} \right)} ~,
\ee
\noindent where $\langle \hat P_0 \rangle$ is the expectation value of the charge in the preferred circle and related to the general result by
\be
\langle P_0 \rangle = \frac{k \bar \ell}{2} + \langle \hat P_0 \rangle~.
\ee

This result yields an interesting consequence. Different limits in the canonical entropy yield different results for the same expectations values of the charges! This casts doubts on the interpretation of the first term as a microcanonical result. While this might seem obvious given that the contribution is imaginary, in cases of interest in holography this contribution might be real  \cite{Detournay:2012pc} and we might wonder whether it can be interpreted as microcanonical counting. In any case, for the case of interest in this work, the Warped Weyl Fermion, we see that charges are quantized in units of $\sqrt{\frac{k}{2}}$. Therefore the naive microcanonical density of states is
\be\label{rhonaive}
\rho_{naive} \sim e^{S} = e^{ 2 \pi \sqrt{ \frac{c}{6} \left( \langle L_0 \rangle - \frac{\langle P_0 \rangle^2}{k} \right)} } e^{2 \pi i n} = e^{ 2 \pi \sqrt{ \frac{c}{6} \left( \langle L_0 \rangle - \frac{\langle P_0 \rangle^2}{k} \right)} } , ~
\ee
\noindent where $n = 2\frac{\hat P^*_0 \langle \hat P_0 \rangle}{k} \in \mathbb{Z}$. Hence, there seems to be no ambiguity in this case.

\subsection{Micro-canonical ensemble}
In order to obtain the micro-canonical density of states from canonical computations we must perform the following integral transform
\be\label{eq:4rho}
\rho(p,h)=\int d\bar \tau \,d\tau \,e^{2\pi i\tau h -2\pi i \bar \tau p}  Z_{\bar \ell | \ell} (\bar \tau |\tau) ~.
\ee
The contour of integration is for both variables over the real axis. 

Let us first consider a simpler problem. Suppose we are interested in calculating the density of states given a value of $L_0$ charge $h$ but summing over all possible $P_0$ charges. Then, we can just plug in $\bar \tau=0$ and compute
\be
\rho(h)=\int  d\tau\, e^{2\pi i\tau h} Z_{\bar \ell | \ell} (0 |\tau)  ~.
\ee
Here we can assume the integral will be dominated at large $h$ by a saddle present at $\tau \rightarrow  - i 0$. We can therefore use the expression
\be
Z_{\bar \ell | \ell} (0 |\tau)  \sim e^{-\frac{i \pi c \ell }{12 \tau}} ~.
\ee
The saddle of the integral if found at $\tau^2 = - \frac{c \ell}{24 h}$, which at large $h$ is consistent with the high temperature limit. We can then evaluate the integral to leading order as
\be
\rho(h) =  e^{2\pi \sqrt{\frac{c}{6} h \ell}} ~.
\ee
Notice this result obviously agrees with our expectations from the canonical ensemble (\ref{rhonaive}). 

Now we would like to go back to the calculation of $\rho(p,h)$. The problem now is that we expect many saddles: one contribution for each charge $\hat P^*_0$ in the spectrum that corresponds to the real part of $\bar \tau$. Going through the same mechanics as above we expect each saddle to contribute, which in the limit $p\gg 1$ and $h - \frac{p^2}{k} \gg 1$ is given by
\be
\rho(p,h)_{\hat P^*_0} =  e^{4 \pi i \frac{ P^*_0}{k} \left( p - \frac{k \bar \ell}{2}\right)} e^{2\pi \sqrt{\frac{c}{6} \left(h \ell - \frac{p^2}{k}\right) } }~.
\ee
One would expect that one needs to sum over all values of $\hat P^*_0$. If they are quantized in units of $\sqrt{\frac{k}{2}}$ like in our Warped Weyl Fermion, as well as the charges $\hat p \equiv p - \frac{k \bar \ell}{2}$, then the sum becomes delta function localized over the physical spectrum and the growth of states with given charge depends only on the spectral flow invariant combination:
\be\label{eq:rph}
\rho(p,h) =  e^{2\pi \sqrt{\frac{c}{6} \left(h \ell - \frac{p^2}{k}\right) } }~.
\ee
In more complicated cases where the spectrum is not distributed as that of the Warped Weyl Fermion or, as in holography, $\hat P^*_0$ becomes imaginary, it is not clear how to interpret microcanonically the term in the canonical entropy depending on the $\bar \tau$ potential.

\section{Discussion}

WCFTs are interesting non-relativistic field theories, with some rather peculiar and novel features. To display some of these peculiarities, we have discussed in length the simplest realization of a WCFT: a massive Weyl fermion in two spacetime dimensions. To summarize, the most striking oddities of our findings are:
\begin{enumerate}
\item In section \ref{sec:2grq} we showed that the angular momentum $J$ is bounded from above. This is intimately related to spectral flow transformations; we are not aware of any relativistic system that has this same property.
\item A WCFT has a preferred frame; one where the spatial cycle is $(\bar\ell,\ell)=(0,1)$. In this frame, modular properties of the torus partition function resemble those of weak Jacobi modular forms, but with some important differences which we will discuss below.
\item The partition function of periodic (Ramond) fermions in a WCFT is non-trivial. Using this new character, we built  two new fermionic partition functions in section \ref{sec:prim};  the spectrum is unitary and the functions are modular covariant.
\item If the spectrum of $P_0$ is real and evenly quantized, which is the case for the Weyl fermion, the canonical and micro-canonical entropy at high temperatures is governed by only the spectral flow invariant combinations. However, if the spectrum of $P_0$ deviates from this case we cannot estimate the density of states universally.
\end{enumerate}

There are some further generalizations of our discussion that are worth highlighting. In section \ref{sec:plane} we have demanded quantization on a particular circle ($\bar \ell = \ell =1$). This is arbitrary and obscures the rescaling invariance of the level $k$ in a $U(1)$ Kac-Moody algebra. The generalization is straight forward and goes as follows: if one wanted to demand angular momentum quantization in an arbitrary circle parameterized by $(\bar \ell, \ell)$,  one is led to the quantization condition (for a theory including the NS sector):
\be\label{eq:xx1}
6\bar \ell^2 k + c = 24 p~,
\ee
\noindent where $c$ is the total central charge and $p$ an arbitrary integer. Notice this agrees with our previous formulae in section \ref{sec:2grq}. Also, this is manifestly dimensionless as $k$ has units of $\bar \ell^{-2}$. As such we see that the only thing that can be fixed is the dimensionless quantity $k \bar \ell^2$, which one can interpret as magnetic flux from the spectral flow perspective. Indeed $\bar \ell^2 k$ is just the $P_0$ charge of the state that corresponds to the vacuum in the canonical circle $(\bar \ell, \ell) =(0,1)$. This is a physical parameter of our theory and plays a role similar to the central charge.

In addition to the quantization condition \eqref{eq:xx1}, we can infer another quantization condition from modular properties of $\hat Z$. Recall that in the canonical circle $(\bar \ell, \ell) =(0,1)$, the $S$ and $T$ transformations are given by \eqref{eq:shz} and  \eqref{eq:thz} respectively. Acting with $S^2$ we find 
\be\label{eq:xx2}
 \hat Z (z |\tau)= \hat Z (-z |\tau)~,
\ee
which gives a reflection symmetry of the spectrum of $P_0$. Acting  with $(ST)^3$ on the partition function gives
\be\label{eq:xx3}
 \hat Z (z |\tau)= e^{i \pi {3k\over 2}}\hat Z (-z |\tau)~.
\ee
Combining \eqref{eq:xx2} and \eqref{eq:xx3} we find that consistency of the modular transformations requires
\be
k\in {4\over 3} n ~,\quad n \in \ZZ~.
\ee
In combination with \eqref{eq:xx1} this implies that any modular invariant WCFTs  must have a central charge $c$ that is a multiple of 8.

WCFTs were originally encountered as the asymptotic symmetry algebra of three dimensional warped black holes  \cite{Compere:2008cv}, and later appeared in more general extremal black holes, see e.g. \cite{Compere:2013bya, Troessaert:2013fma,Compere:2014bia}. These holographic examples do have some peculiarities that are not easy to account for using our fermionic system. For instance, to match the black hole entropy using the thermal properties of a WCFT one needs that $\hat P_0^*$ is purely imaginary \cite{Detournay:2012pc}.  If there is an imaginary contribution to the spectrum of $\hat P_0^*$ it is not clear how to perform the contour integrals in \eqref{eq:4rho}; and in the cases where we can perform the integral the first term in \eqref{eq:4s} drops out. This is an important open question regarding the number of microstates of extremal black holes; gravity typically computes canonical quantities through the holographic dictionary, but it is not evident that these quantities are counting a micro-canonical density of states.

The modular properties of warped partition functions share many similarities with weak Jacobi forms. These modulars forms are those relevant for the counting of dyons in $N=4$ string states, and also crucial to account for the Bekenstein Hawking entropy of BPS black holes \cite{Dijkgraaf:1996it,Dijkgraaf:2000fq,LopesCardoso:2004xf,Jatkar:2005bh,Manschot:2007ha}. In this context, $P_0$ is an R-charge in a supersymmetric algebra, and without loss of generality can be taken to be quantized in units of the $U(1)$ level. If this is the case, the statistical mechanics interpretation of weak Jacobi forms is \cite{Dijkgraaf:2000fq}  
\be
\rho(p,h)=\int_0^1 dz \, e^{-2\pi i z p}  \int_{\epsilon-i\infty}^{\epsilon +i\infty}{d\beta\over 2\pi i} \,e^{2\pi \beta h}  Z (z |\tau)~, \quad \tau=-i\beta~.
\ee
Performing this Laplace transform is straightforward \cite{Kraus:2006nb}, which results in \eqref{eq:rph}. We cannot apply these same assumptions in a WCFT without likely loosing interesting features that relate to extremal non-BPS black holes. Still, it might be worthwhile understanding if some of the elegant mathematical structure behind  weak Jacobi forms could unveil universal properties of the partition functions of WCFTs.

\section*{Acknowledgements}
We are grateful of Francesco Benini, Dalit Engelhardt, P\' eter L\' evay, Gerben Oling, Steve Shenker and Erik Verlinde. A.C. is supported by Nederlandse Organisatie voor Wetenschappelijk Onderzoek (NWO) via a Vidi grant. G.S. would like to thank the Delta-ITP consortium and the Gordon and Betty Moore Foundation for financial support. This work is part of the Delta ITP consortium, a program of the Netherlands Organisation for Scientific Research (NWO) that is funded by the Dutch Ministry of Education, Culture and Science (OCW). This research was supported in part by the National Science Foundation under Grant No. NSF PHY11-25915.


\appendix

\section{Brief summary on WCFTs}\label{app:wcft}
In this appendix we gather some basic properties of WCFTs. The following equations are based on the results in  \cite{Detournay:2012pc}, with the minor caveat that we have adapted some of their expressions to our notation. 

Consider the theory defined on the $(z,w)$ plane as in \eqref{zcoor}-\eqref{wcoor}. On this plane, we denote $T(z)$ as the right moving energy momentum tensor and $P(z)$ a right moving $U(1)$ Kac Moody current. We define
\be
L_n =-{i\over 2\pi}\int dz\, \zeta_n(z) T(z)~,\quad P_n =-{1\over 2\pi}\int dz\, \chi_n(z) P(z)~,
\ee
where we choose the test functions as $\zeta_n=z^{n+1}$ and $\chi_n=z^n$. In terms of the plane charges $(L_n,P_n)$ the commutations relations are
\bea
\label{eq:canonicalgebra}
[ L_n, L_{n'}] &=&(n-n') L_{n+n'}+\frac{c}{12}n(n^2-1)\delta_{n,-n'}~, \cr
[ L_n, P_{n'}] &=&-n'  P_{n'+n}~, \cr
[ P_n, P_{n'}] &=&k \frac{n}{2}\delta_{n,-n'}~,
\eea
which is a Virasoro-Kac-Moody algebra with central charge $c$ and level $k$.

$T(z)$ generates infinitesimal coordinates transformations in $z$, and $P(z)$ generates a gauge transformation in the gauge bundle along $w$. This is the content of the commutation relations \eqref{eq:canonicalgebra}. We can think of these transformations as finite coordinate transformations
\be\label{app1}
w~ \to~ w=w' + g(z')  ~,\quad  z ~\to ~ z=f(z') ~,
\ee
 and in this case, the finite transformation properties are
 \bea\label{app:finite}
P'(z') &=& {\partial z\over \partial z'}\le(P(z)+{k\over 2}{\partial w'\over \partial z} \ri)~,\cr
T'(z') &=& \le({\partial z\over \partial z'}\ri)^2\le(T(z)-{c\over 12}\{z',z\} \ri) + {\partial z\over \partial z'}{\partial w\over \partial z'} P(z) - {k\over 4}\le({\partial w\over \partial z'}\ri)^2~,
\eea
where
\be
\{z',z \}= {{\partial^3 z'\over \partial z^3}\over {\partial z'\over \partial z}} -{3\over 2}\le({{\partial^2 z'\over \partial z^2}\over {\partial z'\over \partial z}}\ri)^2~.
\ee

Among these finite transformations, there is one that is rather interesting. Consider doing a tilt of the $w$ direction:
\be
z=z'~,\quad w= w' + 2\gamma z'~.
\ee
Under this tilt, the currents transform as
\bea
P'(z') &=& P(z)-{k}\gamma~,\cr
T'(z') &=& T(z) -2\gamma P(z) - {k}\gamma^2~,
\eea
which implies that the modes on the plane transform as
\bea
\label{eq:spectralflow}
 L_n &\to & L_n^{(\gamma)}= L_n+2\gamma\,  P_n+{\gamma^2} \,k\, \delta_{n,0}~,\cr
 P_n &\to & P_n^{(\gamma)}=P_n +\gamma \,k\, \delta_{n,0}~.
\eea
This is the usual spectral flow transformation, which leaves the commutations relations \eqref{eq:canonicalgebra} invariant.

Our final remark is regarding unitary representations. The currents $T(z)$ and $P(z)$ are hermitian, and hence
\be\label{eq:pr}
L_{-n}=L^\dagger_n~,\quad P_{-n}=P^\dagger_n~.
\ee
This choice is tied to the sign of $k$, which we assume to be a real positive number. A primary state is defined as a state $|p,h\rangle$ who is an eigenstate of the zero modes
\be
P_0|p,h\rangle = p |p,h\rangle~, \quad L_0|p,h\rangle=h|p,h\rangle~,
\ee
and is annihilated by $(L_{n},P_n)$ with $n>0$. Descendants are created by acting with $L_{-n}$ and $P_{-n}$ ($n>0$). We can derive some simple unitarity bounds on the plane charges by requiring the norm of states to be positive. For instance
\bea
|| L_{-n}|p,h\rangle|| \geq 0 \quad \Rightarrow \quad h\geq 0~,\quad c\geq 0~,\cr
|| P_{-n}|p,h\rangle|| \geq 0 \quad \Rightarrow  \quad p\in \RR ~,\quad  k\geq0~.
\eea
However there are further constraints. Consider the Kac determinant at level one: we have two states created by $P_{-1}$ and $L_{-1}$, and the matrix of inner products is
\be\label{eq:b1}
\left(
\begin{array}{cc}
\langle L_{1} L_{-1} \rangle & \langle P_{1} L_{-1} \rangle \\ 
\langle L_{1} P_{-1} \rangle & \langle P_{1} P_{-1} \rangle
\end{array} \right) = \left(
\begin{array}{cc}
2h & p  \\ 
p & {k\over 2}
\end{array} \right) \quad \Rightarrow \quad h\geq {p^2\over k}~,
\ee
and at level 2 we would find that $c\geq1$. Another way to derive these bounds is by defining a new right moving stress tensor that removes the contributions for the $U(1)$ current. More explicitly, consider
\be\label{eq:lsug}
L'_n=L_n -{1\over k}\sum_{n'} :P_{n+n'}P_{-n'}:~.
\ee
This operator is Hermitian, it commutes with $P_n$ and satisfies 
\be
[ L'_n, L'_{n'}] =(n-n') L'_{n+n'}+\frac{c-1}{12}n(n^2-1)\delta_{n,-n'}~.
\ee
The norm of the states $L_{n}' |p,h\rangle$ will give the same bound as in \eqref{eq:b1}, and that $c\geq1$.  Primary states that saturate the bound, i.e. $h={p^2\over k}$, will be denoted Kac-Moody primaries.


\section{Standard modular forms and functions}\label{app:theta}

Our conventions for theta functions
\bea
\theta_1(z|\t) &=&-i\sum_{r\in \mathbb{Z}+1/2}(-1)^{r-1/2} y^r q^{r^2/2}~, \cr
\theta_2(z|\t) &=&\sum_{r\in \mathbb{Z}+1/2} y^r q^{r^2/2}~, \cr
\theta_3(z|\t) &=&\sum_{n\in \mathbb{Z}} y^n q^{n^2/2}~, \cr
\theta_4(z|\t) &=&\sum_{n\in \mathbb{Z}} (-1)^n y^n q^{n^2/2}~, 
\eea
where $q=e^{2\pi i \t}$ and $y=e^{2\pi i z}$. As a triple product we have
\bea
\theta_1(z|\t) &=&-i y^{1/2} q^{1/8} \prod_{n=1}^\infty (1-q^n)\prod_{n=0}^\infty (1- q^{n+1} y)(1-q^n/y)~, \cr
\theta_2(z|\t) &=&y^{1/2} q^{1/8} \prod_{n=1}^\infty (1-q^n)\prod_{n=0}^\infty (1+ q^{n+1} y)(1+q^n/y)~, \cr
\theta_3(z|\t) &=&\prod_{n=1}^\infty (1-q^n)\prod_{n=0}^\infty (1+ q^{n+1/2} y)(1+q^{n+1/2}/y)~, \cr
\theta_4(z|\t) &=&\prod_{n=1}^\infty (1-q^n)\prod_{n=0}^\infty (1- q^{n+1/2} y)(1-q^{n+1/2}/y)~.
\eea

The quasi-periodicities of the theta functions are
\bea
\label{eq:quasiperiod}
\theta_1(z|\t) &=& (-1)^{a+b} y^b q^{\frac{b^2}{2}}\theta_1(z+a+b \t|\t)~,\cr
\theta_2(z|\t) &=& (-1)^{a} y^b q^{\frac{b^2}{2}}\theta_2(z+a+b \t| \t)~,\cr
\theta_3(z|\t) &=&  y^b q^{\frac{b^2}{2}}\theta_3(z+a+b\t |\t)~,\cr
\theta_4(z|\t) &=& (-1)^{b} y^b q^{\frac{b^2}{2}}\theta_4(z+a+b \t|\t)~.
\eea 
where $a,b\in\ZZ$. The Jacobi identities, needed to perform the $S$ transformation, read as
\bea
\label{eq:jacobi}
\theta_1(z|\t) &=&-i\frac{e^{-i\pi\frac{ z^2}{\t}}}{\sqrt{-i \t}} \theta_1(\frac{z}{\t}|-\frac{1}{\t})~, \cr
\theta_2(z|\t) &=&\frac{e^{-i\pi\frac{ z^2}{\t}}}{\sqrt{-i \t}} \theta_4(\frac{z}{\t}|-\frac{1}{\t})~, \cr
\theta_3(z|\t) &=&\frac{e^{-i\pi\frac{ z^2}{\t}}}{\sqrt{-i \t}} \theta_3(\frac{z}{\t}|-\frac{1}{\t})~, \cr
\theta_4(z|\t) &=&\frac{e^{-i\pi\frac{ z^2}{\t}}}{\sqrt{-i \t}} \theta_2(\frac{z}{\t}|-\frac{1}{\t})~.
\eea
For the $T$ transformation we easily derive from the definition (using that an integer squared has the same parity as itself) that
\bea
\label{eq:Tshifts}
\theta_1(z|\t+1) &=&e^{i\frac{\pi}{4}}\theta_1(z|\t)~, \cr
\theta_2(z|\t+1) &=&e^{i\frac{\pi}{4}}\theta_2(z|\t)~,  \cr
\theta_3(z|\t+1) &=&\theta_4(z|\t)~, \cr
\theta_4(z|\t+1) &=&\theta_3(z|\t)~.
\eea

The Dedekind's eta function is
\be
\eta(\tau)= q^{1/24} \prod_{n=1}^{\infty}(1-q^n)~,\quad q=e^{2\pi i \t}~,
\ee
and its modular properties are
\be\label{eq:etaTS}
\eta(\tau+1)=e^{i\pi/12}\eta(\tau)~,\quad \eta(-1/\tau)=\sqrt{-i\t}\,\eta(\t)~.
\ee


\section{Characters from the fermionic path integral}\label{app:fude} 

In this appendix we derive the partition function of a free Warped Weyl fermion using functional determinants. There are various ambiguities in the regularization of the determinants, which remain unclear to us; in the following we will just apply a regularization scheme that preserves holomorphicity and we discuss how this procedure relates to the results in section \ref{sec:torusweyl}. Similar manipulations for chiral fermions have been done recently in  e.g. \cite{Ryu2012Interacting,Ng:2014sqa}.

The Euclidean action for the free fermion is
\be
I_E= \int dt_Ed\varphi \le(i\bar \Psi\partial_+ \Psi + m\bar\Psi \Psi\ri)~,
\ee
where, with a high abuse of notation, we have $x^+= \varphi +it_E $. This warped conformal theory lives now on a torus, where the identifications are 
\be\label{torid1}
(x^- , x^+) \sim (x^- - 2\pi, x^+ + 2\pi) \sim (x^- - 2\pi \tau, x^+ + 2 \pi \bar{\tau})~.
\ee
However, we could have chosen our cycles rather differently as we discussed in section \ref{sec:ontorus}. And in particular, the preferred frame is that in \eqref{eq:huv} where the identifications are
\be\label{eq:uv3}
(\hat u , \hat v) \sim (\hat u - 2\pi , \hat v ) \sim (\hat u - 2\pi \tau, \hat v + 2 \pi z)~,\quad z \equiv  \bar\tau-\tau~.
\ee
The reason why this frame is preferred in the path integral has to do with our regularization scheme; this will be evident shortly.
The Euclidean path integral is
\be\label{eq:detz}
Z_E= \int D\Psi D\bar\Psi \, e^{-I_E} = \det\le(i\partial_{\hat v} +m\ri) ~.
\ee
The eigenvalue problem we need to solve is
\be
(i\partial_{\hat v} +m)\Psi_{\vec{n}} = \lambda_{\vec{n}} \Psi_{\vec{n}}~,
\ee
and a basis of solutions are exponential functions
\be
\Psi_{\vec{n}} =  \exp( i(m - \lambda_{\vec n}) \hat v + \zeta_{\vec n}\, \hat u)~.
\ee
The eigenvalues depend on two integers -- $\vec n=(n_1,n_2)$, $n_{1,2}\in \Z$ -- since we have to impose on $\Psi_{\vec{n}}$ boundary conditions that are compatible with the topology of the torus. Given \eqref{torid1}, the periods of the eigenfunctions are
\bea\label{eq:bc2}
\Psi_{\vec{n}} (\hat u -2\pi, \hat v)&=& e^{2\pi i(\nu_1 +n_1)}\Psi_{\vec{n}} (\hat u , \hat v)~, \cr
 \Psi_{\vec{n}} (\hat u - 2\pi \tau, \hat v + 2 \pi z)&= &e^{2\pi i(\nu_2 +n_2)}\Psi_{\vec{n}} (\hat u , \hat v)~.
\eea
Here $\nu_{1,2}$ is either zero or one half: this accounts for choosing either R or NS boundary conditions for either the thermal and spatial cycle.  By demanding \eqref{eq:bc2}, we infer that the eigenvalues are 
\bea\label{eq:l3}
\lambda_{\vec{n}}=-{1\over z} \le(n_2+\nu_2-m z -\le(n_1 +\nu_1 \ri)\tau\ri)  ~,
\eea
and there is an similar expression for $\zeta_{\vec{n}}$, which is not important here. The determinant \eqref{eq:detz} now reads
\be\label{eq:11}
 \det(i\partial_{\hat v} +m)=\prod_{n_1,n_2=-\infty}^\infty \lambda_{\vec n} ~.
\ee

Let us first discuss anti-periodic (NS) boundary conditions on both cycles, i.e. $\nu_1=\nu_2=1/2$. To regulate the product we will make use of two identities:
\be\label{eq:1}
\prod_{n=-\infty}^{\infty} \left(n+{1\over 2}+ a\ri) = -2i \cos(\pi a)~,
\ee 
and
\be\label{eq:2}
\prod_{n=0}^\infty q^{-{1\over 2}(n+{1\over2}\pm a)^{-s}}= q^{-{1\over 2}\zeta(s,1/2\pm a)}~,
\ee
where $\zeta(s,x)$ is the Hurwitz zeta function and, in particular, we will use
\be\label{eq:3}
\zeta(-1,1/2\pm a)= {1\over 12} -{a^2\over 2}~.
\ee
These formulas do contain some ambiguities, which we will comment on below. Using these identities, we find
\bea\label{eq:3}
\prod_{n_1,n_2=-\infty}^\infty \le(n_2+{1\over 2}-m z -\le(n_1 +{1\over 2} \ri)\tau\ri)
= q^{-{1\over 24}}  \prod_{n=0}^\infty \left(1+ y^{-1} \,q^{n+{1\over 2}}\right)  \left(1+y\,q^{n+{1\over 2}}\right)
\eea
where 
\be
q\equiv  e^{-2\pi i\tau}~,\quad y\equiv e^{2\pi i m z}~.
\ee
And hence the path integral, up to an overall normalization is
\be\label{eq:zens}
Z_E(z|\tau)_{{1\over 2},{1\over2}} = \hat Z_{\rm NSNS}(z|\tau)
\ee
which is the result reported in \eqref{eq:zis}. 

The regularization scheme we chose in \eqref{eq:1}-\eqref{eq:3} is such that we obtain exactly the NS sector partition function. However, we could have chosen a different scheme. For instance, in  \cite{Quine1993Zeta} their regularization procedure gives the following identity 
\be\label{eq:4}
\prod_{n_1,n_2=-\infty}^{\infty} \left(n_2+n_1 \tau + z\ri) = i\eta(\tau)^{-1}\exp\le(-{\pi i\tau\over 6}-\pi i z\ri) \theta_1(z|\tau) ~,\quad\quad {\rm Im}\tau>0~.
\ee 
This scheme preserves invariance of $Z_E(z| \tau)$ under $T$ -- $(\tau,z) \to(\tau +1,z)$ -- but it would differ from $\hat Z(z|\tau)$ which is anomalous under $T$. There are at least two concrete reasons for the ambiguities. First, the product we are considering in \eqref{eq:11} involves complex numbers and hence there is no unambigous ordering of the eigenvalues $\lambda_{\vec n}$. The second reason could be attributed to the double product; this makes ambiguous how to identify $z$ in \eqref{eq:3}. See \cite{Quine1993Zeta} and references within for further discussion. 

Our scheme also favors the coordinates \eqref{eq:uv3} since in both   \eqref{eq:1}-\eqref{eq:3} and \eqref{eq:4} we treat $z$ and $\tau$ as independent variables. In comparison, for the coordinates \eqref{torid1} we should consider a scheme that treats $\tau$ and $\bar \tau$ independently. The difference between these two frames is accounted by a spectral flow transformation \eqref{spectral}; this introduces anomalies which are more subtle to capture in the measure of the path integral. 

From \eqref{eq:zens} we can obtain all the remaining fermionic sectors by spectral flow and a $S$ transformation. In particular, shifting $z$ effectively changes the spatial boundary conditions; for example
\be\label{eq:zcs}
 \hat Z_{\rm R NS}(z|\tau)= q^{1/8} y^{1/2}\hat Z_{\rm NSNS}(z-{\tau \over 2}|\tau)~.
\ee
This should be viewed as a passive  spectral flow transformation, since the shift in $z$ is interpreted as flow between NS and R while keeping the coordinates on the torus fixed. The anomalous contributions in \eqref{eq:zcs} account for the change in vacuum energies for each sector, and are accounted by the anomalous piece of \eqref{eq:spectralflow} with $\gamma={1\over 4}$. 

%


\bibliographystyle{utphys}
\bibliography{wcft}

\providecommand{\href}[2]{#2}\begingroup\raggedright\begin{thebibliography}{10}

\bibitem{Migdal:1972tk}
A.~A. Migdal, ``{Conformal invariance and bootstrap},''
\href{http://dx.doi.org/10.1016/0370-2693(71)90211-5}{{\em Phys. Lett.}
  {\bfseries B37} (1971) 386--388}.

\bibitem{Polyakov:1974gs}
A.~M. Polyakov, ``{Nonhamiltonian approach to conformal quantum field
  theory},''
{\em Zh. Eksp. Teor. Fiz.} {\bfseries 66} (1974) 23--42.

\bibitem{Ferrara:1973yt}
S.~Ferrara, A.~F. Grillo, and R.~Gatto, ``{Tensor representations of conformal
  algebra and conformally covariant operator product expansion},''
\href{http://dx.doi.org/10.1016/0003-4916(73)90446-6}{{\em Annals Phys.}
  {\bfseries 76} (1973) 161--188}.

\bibitem{Rattazzi:2008pe}
R.~Rattazzi, V.~S. Rychkov, E.~Tonni, and A.~Vichi, ``{Bounding scalar operator
  dimensions in 4D CFT},''
  \href{http://dx.doi.org/10.1088/1126-6708/2008/12/031}{{\em JHEP} {\bfseries
  12} (2008) 031},
\href{http://arxiv.org/abs/0807.0004}{{\ttfamily arXiv:0807.0004 [hep-th]}}.

\bibitem{Witten:2007ct}
E.~Witten, ``{Conformal Field Theory In Four And Six Dimensions},'' in {\em
  {Topology, geometry and quantum field theory. Proceedings, Symposium in the
  honour of the 60th birthday of Graeme Segal, Oxford, UK, June 24-29, 2002}}.
\newblock 2007.
\newblock \href{http://arxiv.org/abs/0712.0157}{{\ttfamily arXiv:0712.0157
  [math.RT]}}.
\newblock
\url{http://inspirehep.net/record/769452/files/arXiv:0712.0157.pdf}.
\newblock

\bibitem{Ginsparg:1988ui}
P.~H. Ginsparg, ``{APPLIED CONFORMAL FIELD THEORY},'' in {\em {Les Houches
  Summer School in Theoretical Physics: Fields, Strings, Critical Phenomena Les
  Houches, France, June 28-August 5, 1988}}.
\newblock 1988.
\newblock \href{http://arxiv.org/abs/hep-th/9108028}{{\ttfamily
  arXiv:hep-th/9108028 [hep-th]}}.
\newblock
\url{http://inspirehep.net/record/265020/files/arXiv:hep-th_9108028.pdf}.
\newblock

\bibitem{yellowpages}
P.~Mathieu, D.~Senechal, and {P. di Francesco}, ``{C}onformal {F}ield
  {T}heory,'' {\em Springer} (1997) .

\bibitem{Cardy:1986ie}
J.~L. Cardy, ``{Operator Content of Two-Dimensional Conformally Invariant
  Theories},''
\href{http://dx.doi.org/10.1016/0550-3213(86)90552-3}{{\em Nucl.Phys.}
  {\bfseries B270} (1986) 186--204}.

\bibitem{DiPietro:2014bca}
L.~Di~Pietro and Z.~Komargodski, ``{Cardy formulae for SUSY theories in $d =$ 4
  and $d =$ 6},'' \href{http://dx.doi.org/10.1007/JHEP12(2014)031}{{\em JHEP}
  {\bfseries 12} (2014) 031},
\href{http://arxiv.org/abs/1407.6061}{{\ttfamily arXiv:1407.6061 [hep-th]}}.

\bibitem{Shaghoulian:2015kta}
E.~Shaghoulian, ``{Modular forms and a generalized Cardy formula in higher
  dimensions},''
\href{http://arxiv.org/abs/1508.02728}{{\ttfamily arXiv:1508.02728 [hep-th]}}.

\bibitem{Basar:2015xda}
G.~Basar, A.~Cherman, K.~R. Dienes, and D.~A. McGady, ``{A 4D-2D equivalence
  for large-N Yang-Mills theory},''
\href{http://arxiv.org/abs/1507.08666}{{\ttfamily arXiv:1507.08666 [hep-th]}}.

\bibitem{Gonzalez:2011nz}
H.~A. Gonzalez, D.~Tempo, and R.~Troncoso, ``{Field theories with anisotropic
  scaling in 2D, solitons and the microscopic entropy of asymptotically
  Lifshitz black holes},''
  \href{http://dx.doi.org/10.1007/JHEP11(2011)066}{{\em JHEP} {\bfseries 11}
  (2011) 066},
\href{http://arxiv.org/abs/1107.3647}{{\ttfamily arXiv:1107.3647 [hep-th]}}.

\bibitem{Shaghoulian:2015dwa}
E.~Shaghoulian, ``{A Cardy formula for holographic hyperscaling-violating
  theories},''
\href{http://arxiv.org/abs/1504.02094}{{\ttfamily arXiv:1504.02094 [hep-th]}}.

\bibitem{Maldacena:1997re}
J.~M. Maldacena, ``{The large N limit of superconformal field theories and
  supergravity},'' {\em Adv. Theor. Math. Phys.} {\bfseries 2} (1998) 231--252,
\href{http://arxiv.org/abs/hep-th/9711200}{{\ttfamily arXiv:hep-th/9711200}}.

\bibitem{Witten:1998qj}
E.~Witten, ``{Anti-de Sitter space and holography},'' {\em Adv. Theor. Math.
  Phys.} {\bfseries 2} (1998) 253--291,
\href{http://arxiv.org/abs/hep-th/9802150}{{\ttfamily arXiv:hep-th/9802150}}.

\bibitem{Gubser:1998bc}
S.~S. Gubser, I.~R. Klebanov, and A.~M. Polyakov, ``{Gauge theory correlators
  from non-critical string theory},''
  \href{http://dx.doi.org/10.1016/S0370-2693(98)00377-3}{{\em Phys. Lett.}
  {\bfseries B428} (1998) 105--114},
\href{http://arxiv.org/abs/hep-th/9802109}{{\ttfamily arXiv:hep-th/9802109}}.

\bibitem{Guica:2008mu}
M.~Guica, T.~Hartman, W.~Song, and A.~Strominger, ``{The Kerr/CFT
  Correspondence},''
\href{http://arxiv.org/abs/0809.4266}{{\ttfamily arXiv:0809.4266 [hep-th]}}.

\bibitem{Castro:2009jf}
A.~Castro and F.~Larsen, ``{Near Extremal Kerr Entropy from AdS$_2$ Quantum
  Gravity},'' \href{http://dx.doi.org/10.1088/1126-6708/2009/12/037}{{\em JHEP}
  {\bfseries 12} (2009) 037},
\href{http://arxiv.org/abs/0908.1121}{{\ttfamily arXiv:0908.1121 [hep-th]}}.

\bibitem{Anninos:2008fx}
D.~Anninos, W.~Li, M.~Padi, W.~Song, and A.~Strominger, ``{Warped AdS$_3$ Black
  Holes},'' \href{http://dx.doi.org/10.1088/1126-6708/2009/03/130}{{\em JHEP}
  {\bfseries 03} (2009) 130},
\href{http://arxiv.org/abs/0807.3040}{{\ttfamily arXiv:0807.3040 [hep-th]}}.

\bibitem{Anninos:2013nja}
D.~Anninos, J.~Samani, and E.~Shaghoulian, ``{Warped Entanglement Entropy},''
  \href{http://dx.doi.org/10.1007/JHEP02(2014)118}{{\em JHEP} {\bfseries 02}
  (2014) 118},
\href{http://arxiv.org/abs/1309.2579}{{\ttfamily arXiv:1309.2579 [hep-th]}}.

\bibitem{Detournay:2012pc}
S.~Detournay, T.~Hartman, and D.~M. Hofman, ``{Warped Conformal Field
  Theory},'' \href{http://dx.doi.org/10.1103/PhysRevD.86.124018}{{\em
  Phys.Rev.} {\bfseries D86} (2012) 124018},
\href{http://arxiv.org/abs/1210.0539}{{\ttfamily arXiv:1210.0539 [hep-th]}}.

\bibitem{Hofman:2011zj}
D.~M. Hofman and A.~Strominger, ``{Chiral Scale and Conformal Invariance in 2D
  Quantum Field Theory},''
  \href{http://dx.doi.org/10.1103/PhysRevLett.107.161601}{{\em Phys.Rev.Lett.}
  {\bfseries 107} (2011) 161601},
\href{http://arxiv.org/abs/1107.2917}{{\ttfamily arXiv:1107.2917 [hep-th]}}.

\bibitem{Hofman:2014loa}
D.~M. Hofman and B.~Rollier, ``{Warped Conformal Field Theory as Lower Spin
  Gravity},'' \href{http://dx.doi.org/10.1016/j.nuclphysb.2015.05.011}{{\em
  Nucl.Phys.} {\bfseries B897} (2015) 1--38},
\href{http://arxiv.org/abs/1411.0672}{{\ttfamily arXiv:1411.0672 [hep-th]}}.

\bibitem{Cardy:1992tq}
J.~L. Cardy, ``{Critical exponents of the chiral Potts model from conformal
  field theory},'' \href{http://dx.doi.org/10.1016/0550-3213(93)90353-Q}{{\em
  Nucl. Phys.} {\bfseries B389} (1993) 577--586},
\href{http://arxiv.org/abs/hep-th/9210002}{{\ttfamily arXiv:hep-th/9210002
  [hep-th]}}.

\bibitem{Cappelli:1996np}
A.~Cappelli and G.~R. Zemba, ``{Modular invariant partition functions in the
  quantum Hall effect},''
  \href{http://dx.doi.org/10.1016/S0550-3213(97)00110-7}{{\em Nucl. Phys.}
  {\bfseries B490} (1997) 595--632},
\href{http://arxiv.org/abs/hep-th/9605127}{{\ttfamily arXiv:hep-th/9605127
  [hep-th]}}.

\bibitem{Cappelli:2010be}
A.~Cappelli, G.~Viola, and G.~R. Zemba, ``{Chiral partition functions of
  quantum Hall droplets},''
  \href{http://dx.doi.org/10.1016/j.aop.2009.10.007}{{\em Annals Phys.}
  {\bfseries 325} (2010) 465--490},
\href{http://arxiv.org/abs/0909.3588}{{\ttfamily arXiv:0909.3588
  [cond-mat.mes-hall]}}.

\bibitem{Ryu:2012he}
S.~Ryu and S.-C. Zhang, ``{Interacting topological phases and modular
  invariance},'' \href{http://dx.doi.org/10.1103/PhysRevB.85.245132}{{\em Phys.
  Rev.} {\bfseries B85} (2012) 245132},
\href{http://arxiv.org/abs/1202.4484}{{\ttfamily arXiv:1202.4484
  [cond-mat.str-el]}}.

\bibitem{Gaberdiel:2012yb}
M.~R. Gaberdiel, T.~Hartman, and K.~Jin, ``{Higher Spin Black Holes from
  CFT},'' \href{http://dx.doi.org/10.1007/JHEP04(2012)103}{{\em JHEP}
  {\bfseries 1204} (2012) 103},
\href{http://arxiv.org/abs/1203.0015}{{\ttfamily arXiv:1203.0015 [hep-th]}}.

\bibitem{deBoer:2013gz}
J.~de~Boer and J.~I. Jottar, ``{Thermodynamics of higher spin black holes in
  $AdS_3$},'' \href{http://dx.doi.org/10.1007/JHEP01(2014)023}{{\em JHEP}
  {\bfseries 1401} (2014) 023},
\href{http://arxiv.org/abs/1302.0816}{{\ttfamily arXiv:1302.0816 [hep-th]}}.

\bibitem{deBoer:2014fra}
J.~de~Boer and J.~I. Jottar, ``{Boundary Conditions and Partition Functions in
  Higher Spin AdS$_3$/CFT$_2$},''
\href{http://arxiv.org/abs/1407.3844}{{\ttfamily arXiv:1407.3844 [hep-th]}}.

\bibitem{Ryu2012Interacting}
S.~Ryu and S.-C. Zhang, ``{Interacting topological phases and modular
  invariance},'' \href{http://dx.doi.org/10.1103/physrevb.85.245132}{{\em
  Physical Review B} {\bfseries 85} no.~24, (June, 2012) 245132},
  \href{http://arxiv.org/abs/1202.4484}{{\ttfamily 1202.4484}}.
  \url{http://dx.doi.org/10.1103/physrevb.85.245132}.

\bibitem{Carlip:1998qw}
S.~Carlip, ``{What we don't know about BTZ black hole entropy},''
  \href{http://dx.doi.org/10.1088/0264-9381/15/11/020}{{\em Class.Quant.Grav.}
  {\bfseries 15} (1998) 3609--3625},
\href{http://arxiv.org/abs/hep-th/9806026}{{\ttfamily arXiv:hep-th/9806026
  [hep-th]}}.

\bibitem{Compere:2008cv}
G.~Compere and S.~Detournay, ``{Semi-classical central charge in topologically
  massive gravity},'' \href{http://dx.doi.org/10.1088/0264-9381/26/1/012001,
  10.1088/0264-9381/26/13/139801}{{\em Class.Quant.Grav.} {\bfseries 26} (2009)
  012001},
\href{http://arxiv.org/abs/0808.1911}{{\ttfamily arXiv:0808.1911 [hep-th]}}.

\bibitem{Compere:2013bya}
G.~Compère, W.~Song, and A.~Strominger, ``{New Boundary Conditions for AdS3},''
  \href{http://dx.doi.org/10.1007/JHEP05(2013)152}{{\em JHEP} {\bfseries 05}
  (2013) 152},
\href{http://arxiv.org/abs/1303.2662}{{\ttfamily arXiv:1303.2662 [hep-th]}}.

\bibitem{Troessaert:2013fma}
C.~Troessaert, ``{Enhanced asymptotic symmetry algebra of $AdS$$\_{3}$},''
  \href{http://dx.doi.org/10.1007/JHEP08(2013)044}{{\em JHEP} {\bfseries 08}
  (2013) 044},
\href{http://arxiv.org/abs/1303.3296}{{\ttfamily arXiv:1303.3296 [hep-th]}}.

\bibitem{Compere:2014bia}
G.~Compère, M.~Guica, and M.~J. Rodriguez, ``{Two Virasoro symmetries in
  stringy warped AdS$\_{3}$},''
  \href{http://dx.doi.org/10.1007/JHEP12(2014)012}{{\em JHEP} {\bfseries 12}
  (2014) 012},
\href{http://arxiv.org/abs/1407.7871}{{\ttfamily arXiv:1407.7871 [hep-th]}}.

\bibitem{Dijkgraaf:1996it}
R.~Dijkgraaf, E.~P. Verlinde, and H.~L. Verlinde, ``Counting dyons in {N = 4}
  string theory,'' {\em Nucl. Phys.} {\bfseries B484} (1997) 543--561,
\href{http://arxiv.org/abs/hep-th/9607026}{{\ttfamily hep-th/9607026}}.

\bibitem{Dijkgraaf:2000fq}
R.~Dijkgraaf, J.~M. Maldacena, G.~W. Moore, and E.~P. Verlinde, ``{A black hole
  farey tail},''
\href{http://arxiv.org/abs/hep-th/0005003}{{\ttfamily arXiv:hep-th/0005003}}.

\bibitem{LopesCardoso:2004xf}
G.~Lopes~Cardoso, B.~de~Wit, J.~Kappeli, and T.~Mohaupt, ``{Asymptotic
  degeneracy of dyonic N = 4 string states and black hole entropy},''
  \href{http://dx.doi.org/10.1088/1126-6708/2004/12/075}{{\em JHEP} {\bfseries
  12} (2004) 075},
\href{http://arxiv.org/abs/hep-th/0412287}{{\ttfamily arXiv:hep-th/0412287
  [hep-th]}}.

\bibitem{Jatkar:2005bh}
D.~P. Jatkar and A.~Sen, ``{Dyon spectrum in CHL models},''
  \href{http://dx.doi.org/10.1088/1126-6708/2006/04/018}{{\em JHEP} {\bfseries
  04} (2006) 018},
\href{http://arxiv.org/abs/hep-th/0510147}{{\ttfamily arXiv:hep-th/0510147
  [hep-th]}}.

\bibitem{Manschot:2007ha}
J.~Manschot and G.~W. Moore, ``{A Modern Farey Tail},''
  \href{http://dx.doi.org/10.4310/CNTP.2010.v4.n1.a3}{{\em Commun. Num. Theor.
  Phys.} {\bfseries 4} (2010) 103--159},
\href{http://arxiv.org/abs/0712.0573}{{\ttfamily arXiv:0712.0573 [hep-th]}}.

\bibitem{Kraus:2006nb}
P.~Kraus and F.~Larsen, ``{Partition functions and elliptic genera from
  supergravity},'' \href{http://dx.doi.org/10.1088/1126-6708/2007/01/002}{{\em
  JHEP} {\bfseries 0701} (2007) 002},
\href{http://arxiv.org/abs/hep-th/0607138}{{\ttfamily arXiv:hep-th/0607138
  [hep-th]}}.

\bibitem{Ng:2014sqa}
G.~S. Ng and P.~Surowka, ``{One-loop effective actions and 2D hydrodynamics
  with anomalies},''
  \href{http://dx.doi.org/10.1016/j.physletb.2015.05.011}{{\em Phys.Lett.}
  {\bfseries B746} (2015) 281--284},
\href{http://arxiv.org/abs/1411.7989}{{\ttfamily arXiv:1411.7989 [hep-th]}}.

\bibitem{Quine1993Zeta}
J.~R. Quine, S.~H. Heydari, and R.~Y. Song, ``{Zeta regularized products},''
  \href{http://dx.doi.org/10.1090/s0002-9947-1993-1100699-1}{{\em Transactions
  of the American Mathematical Society} {\bfseries 338} no.~1, (1993)
  213--231}. \url{http://dx.doi.org/10.1090/s0002-9947-1993-1100699-1}.

\end{thebibliography}\endgroup

\end{document}